\documentclass[12pt]{article}
\usepackage[english]{babel}
\usepackage{graphicx,color}
\usepackage{framed}
\usepackage[normalem]{ulem}
\usepackage{amsmath}
\usepackage{amsthm}
\usepackage{amssymb}
\usepackage{amsfonts}
\usepackage{enumerate}
\usepackage{algorithm}
\usepackage{algpseudocode}
\usepackage[top=1 in,bottom=1in, left=0.8 in, right=0.8 in]{geometry}
\usepackage{bm}
\usepackage{natbib}
\usepackage{multirow}
\usepackage{makecell}
\usepackage{comment}

\theoremstyle{definition}
\newtheorem{Theorem}{\textbf{Theorem}}

\newtheorem{D}{Definition}

\newtheorem{Remark}{Remark}

\newtheorem{asum}{Assumption}

\newcommand{\argmin}{\mathop{\rm arg\min}}

\newcommand{\supp}{{\rm supp}}

\usepackage{xcolor}

\def\bm{\boldsymbol}
\usepackage{setspace}

\AtBeginDocument{%
  \doublespacing
  \setlength{\textheight}{\dimexpr 27\baselineskip+\topskip\relax}
}

\usepackage{hyperref}
\hypersetup{
    colorlinks=true,
    linkcolor=black,
    filecolor=magenta,      
    urlcolor=cyan,
    citecolor = blue,
    pdftitle={Overleaf Example},
    pdfpagemode=FullScreen,
    }

\usepackage {xr}
\makeatletter
\newcommand*{\addFileDependency}[1]{
  \typeout{(#1)}
  \@addtofilelist{#1}
  \IfFileExists{#1}{}{\typeout{No file #1.}}
}
\makeatother

\newcommand*{\myexternaldocument}[1]{
    \externaldocument{#1}
    \addFileDependency{main.tex}
}

\myexternaldocument{SM}
\usepackage{etoolbox}
\AtBeginEnvironment{thebibliography}{
    \small
    \singlespacing
}

\usepackage{titlesec}
\titleformat{\section}
  {\Large\bfseries}
  {\thesection}
  {0.5em}
  {}

\titlespacing{\section}
  {0pt}
  {5pt}
  {0pt}

\titleformat{\subsection}
  {\bfseries\large}
  {\thesubsection}
  {0.5em}
  {}

\titlespacing{\subsection}
  {0pt}
  {0pt}
  {0pt}

\titlespacing*{\paragraph}
  {0pt}
  {0pt}
  {1em}

\AtBeginEnvironment{equation}{%
  \setlength{\abovedisplayskip}{3pt}
  \setlength{\belowdisplayskip}{3pt}
}
\AtBeginEnvironment{equation*}{%
  \setlength{\abovedisplayskip}{3pt}
  \setlength{\belowdisplayskip}{3pt}
}

\AtBeginEnvironment{eqnarray}{%
  \setlength{\abovedisplayskip}{3pt}
  \setlength{\belowdisplayskip}{3pt}
}
\AtBeginEnvironment{equation*}{%
  \setlength{\abovedisplayskip}{3pt}
  \setlength{\belowdisplayskip}{3pt}
}

\title{Panel Flow Matching: A Generative Approach to Learning Distributions of Longitudinal Data}
\date{}
  \author{Jianbin Tan$^1$, ~ 
    Pixu Shi$^1$, ~ and ~
    Anru R. Zhang$^{1,2,}$\thanks{Corresponding author. E-mail: anru.zhang@duke.edu}}

\begin{document}

\maketitle

\footnotetext[1]{Department of Biostatistics \& Bioinformatics, Duke University, Durham, NC, USA}

\footnotetext[2]{Department of Computer Science, Duke University, Durham, NC, USA}

\begin{abstract}
\singlespacing
Learning distributions of longitudinal data is central to tasks such as visualization, completion, classification, and synthetic data generation, but remains statistically challenging because longitudinal observations are often irregular, sparse, and collected from only a limited number of subjects. To address this, we develop a novel generative framework, termed panel flow matching (PFM), for learning longitudinal distributions by pooling information across time via a continuous panel flow model. PFM combines a forward flow-matching step with a backward kernel-fitting step, yielding a flexible and data-adaptive approach for capturing complex distributional structure. We apply PFM to estimate panel densities, namely the cross-sectional densities of longitudinal data, and establish statistical guarantees under irregular and sparse sampling designs. Under this, PFM naturally supports tasks including longitudinal completion, synthetic data generation, and classification, without requiring a preliminary dimension-reduction step to handle data irregularity. Extensive simulations demonstrate that PFM outperforms existing methods across these tasks. We further apply PFM to a vaginal microbiome longitudinal dataset from 188 pregnancies labeled as term or preterm, where it improves classification accuracy and reveals time-varying distributional differences between the two groups.
\end{abstract}

\noindent%

\begin{sloppypar}

\section{Introduction}\label{sec:intro}
Longitudinal data have become increasingly prevalent in real-world applications, including microbiome studies \citep{costello2022vaginal,tan2026associating}, biomedicine \citep{schussler2019longitudinal,johnson2020mimic}, public health \citep{johnson2020mimic,gai2026subtype}, and environmental surveillance \citep{yan2025deep,wang2023regionalization}. For a broad range of longitudinal tasks, understanding the underlying data distribution is essential. 
For example, in data visualization, distribution estimation helps reveal time-varying patterns in probability mass and concentration; in generative modeling and data completion, distributional learning is central to generating or imputing realistic data; and for classification, distributional characteristics are fundamental for capturing heterogeneity and temporal variation.
However, learning distributions from longitudinal data remains statistically challenging, because such data are often available only for a limited number of subjects, while measurements are typically collected at unaligned time points with uneven visit schedules and sparse sampling \citep{costello2022vaginal,johnson2020mimic,zhang2024trajectory,zhang2024individualized,gai2026subtype,tan2025integrated}. This structure makes standard distribution-estimation methods difficult to apply.

In the literature, a common way to handle irregular longitudinal data is through functional data approaches, which use linear dimension reduction, such as functional principal component analysis \citep[FPCA;][]{yao2005functional,hsing2015theoretical,xiao2018fast}, to transform irregular trajectories into structured score representations. Many longitudinal tasks have been developed through FPCA, including classification, data completion, and clustering \citep{hsing2015theoretical,kraus2015components,wang2016functional,xue2024optimal}. However, these methods are primarily driven by mean and covariance structure and often rely on Gaussian assumptions to handle sparse data \citep{yao2005functional,xiao2018fast}. As a result, they are less well suited to longitudinal data with substantial non-Gaussian distributional features.

To accommodate richer distributional structure, a classical alternative is kernel density estimation \citep{takezawa2005introduction,jiang2017uniform,chen2017tutorial}, which estimates distributions by borrowing information from nearby samples through a kernel structure. However, these approaches are mainly developed for structured tabular data, and the neighborhood is defined through a prespecified kernel metric, which may be restrictive for capturing complex distributional features in irregular longitudinal settings.
More recently, a growing body of work has focused on learning distributions through generative modeling, including variational autoencoders \citep[VAE;][]{doersch2016tutorial,ramchandran2021longitudinal}, normalizing flows \citep{papamakarios2021normalizing}, diffusion models \citep{ho2020denoising,yang2023diffusion}, and more general flow-based methods \citep{chen2018neural,liu2022flow,lipman2022flow,tong2023improving,gao2024convergence,tan2025smooth}. These approaches learn generators that transport latent samples to the data space. Because the transport is nonlinear and typically learned by neural networks, these methods offer substantially greater flexibility than classical linear, Gaussian, or kernel-based methods for modeling complex patterns.

More broadly, existing latent generative methods may be viewed through two common perspectives: direct transport and continuous transport. In direct-transport approaches, generation is carried out by mapping a latent variable to the observed space through a single transformation step, as in variational autoencoders and normalizing flows \citep{ramchandran2021longitudinal,papamakarios2021normalizing}.
In continuous-transport approaches, the target distribution is reached through an evolution process indexed by an artificial time variable, as in continuous normalizing flows, diffusion models, and related flow-matching methods \citep{chen2018neural,ho2020denoising,liu2022flow,lipman2022flow}. Rather than specifying the full transformation at once, they construct the map progressively through a sequence of local updates. This formulation can provide additional flexibility for capturing complex nonlinear structure, especially when the target distribution is difficult to represent by a simpler direct transformation.

Despite their success, existing continuous generative methods \citep{ho2020denoising,liu2022flow,lipman2022flow} are mostly developed for vector-valued observations and are generally most suitable for longitudinal data recorded on a common time grid.
Some extensions introduce alternative transport structures to accommodate irregular sampling \citep{zhang2024trajectory,tan2025smooth}. However, they are either primarily developed for longitudinal prediction \citep{zhang2024trajectory} or for one-dimensional longitudinal data generation \citep{tan2025smooth}, rather than for learning the distribution of multivariate longitudinal trajectories.
More importantly, most existing methods rely on complex neural network architectures to model continuous transport over full longitudinal trajectories, which in turn requires large subject sample sizes and densely observed data.
These constraints limit their applicability in settings with limited subject samples and sparse, irregular observations.

\begin{figure}[h]
    \centering
    \includegraphics[scale = 0.4]{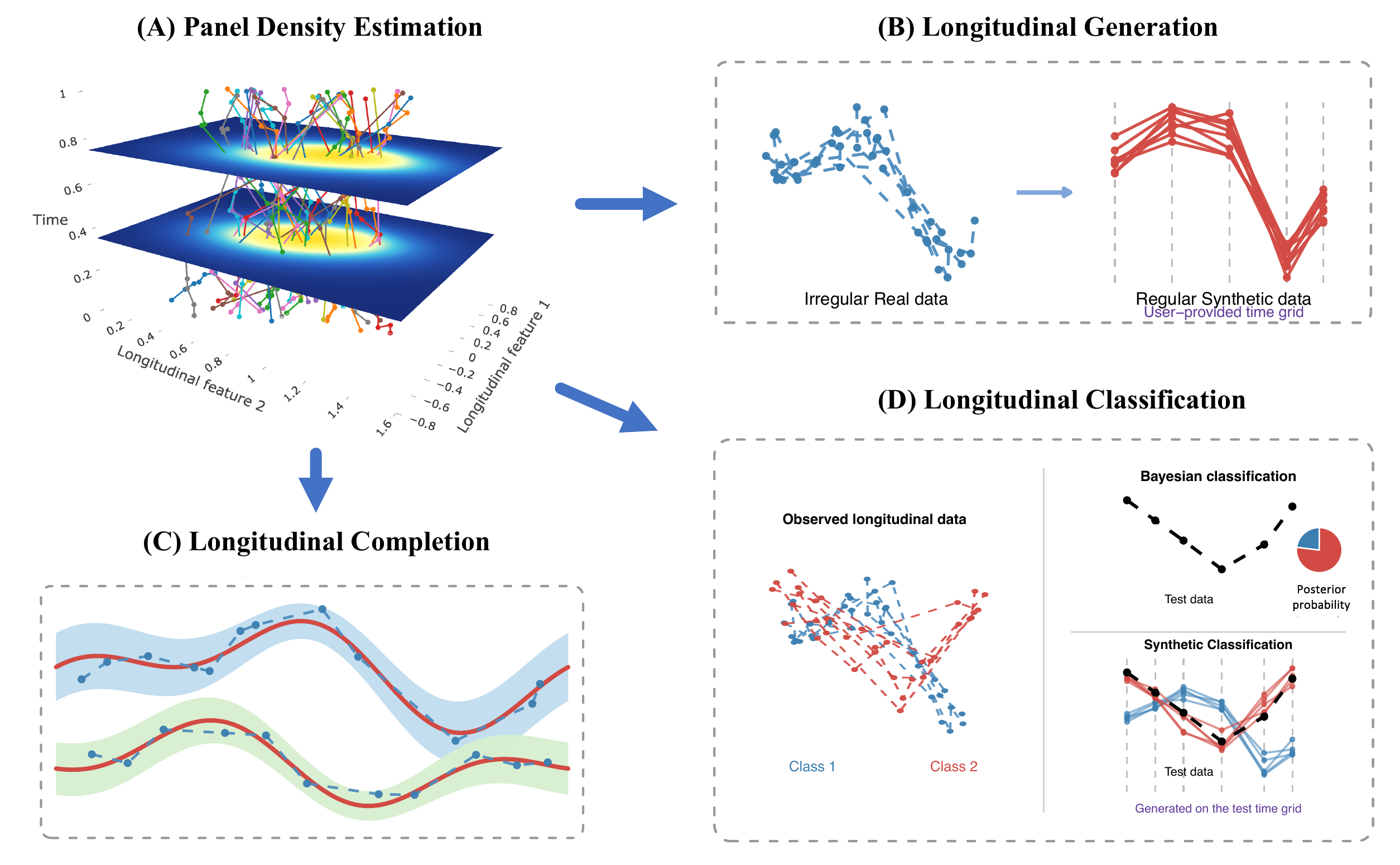}
    \caption{
        \textbf{(A)} A 3D visualization of sparse and irregular two-dimensional longitudinal trajectories together with
cross-sectional density estimates at two selected time points. Individual subject
trajectories are displayed as dashed curves with markers, while
density surfaces highlight regions of high concentration (yellow) and low concentration
(blue) in the $(\text{feature }1,\ \text{feature }2)$ plane. 
\textbf{(B)} Illustration of regular longitudinal generation from irregular data. 
\textbf{(C)} Illustration of longitudinal completion regions from irregular data. 
\textbf{(D)} Illustration of longitudinal classification via Bayesian and synthetic viewpoints.}
    \label{fig: pan_illu}
\end{figure}

To address these challenges, we develop a novel method called {\it panel flow matching} (PFM), a neural-network-based continuous generative framework designed for practical longitudinal data with irregular and sparse sampling and limited subject sample sizes. Our method extends simulation-free flow-matching methods \citep{liu2022flow,lipman2022flow}, with the key distinction that it relies on a novel panelwise flow structure for continuous transport from a latent distribution to time-indexed observed distributions. This construction includes a forward flow-matching step and a backward kernel-fitting step, facilitating estimation of both the continuous transport and the latent distribution from irregular and sparse data. It also has substantially lower model complexity than existing trajectory-level flow-based methods, and its estimation borrows strength across both subjects and time, reducing the reliance on large datasets.

\begin{figure}[h]
    \centering
    \includegraphics[width=0.9\linewidth]{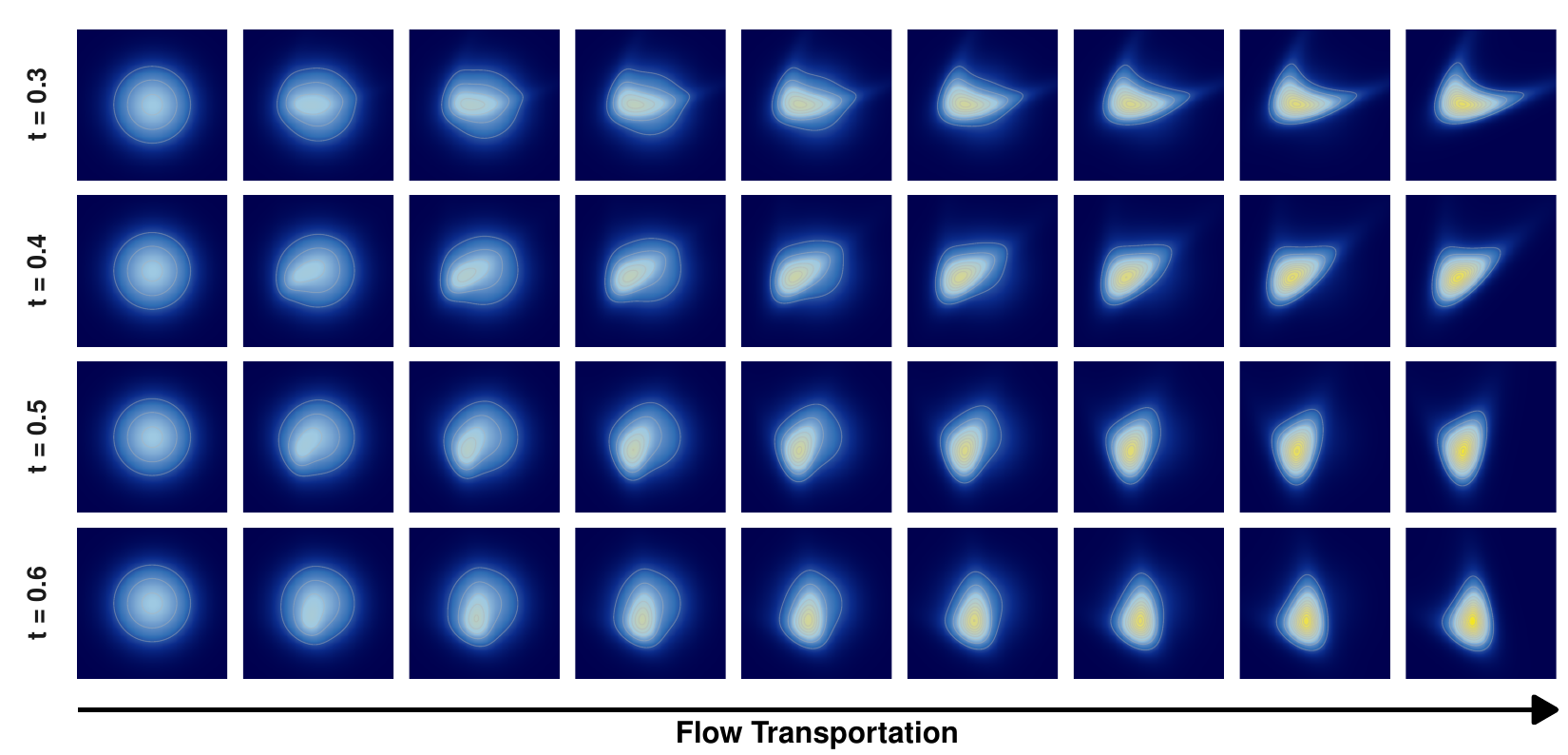}
    \caption{Flow evolution of two-dimensional panel densities $p_t$ over longitudinal time $t$.
    The panels illustrate the transport from a Gaussian latent distribution to the target densities $p_t$ at four time points ($t=0.3,0.4,0.5,0.6$).}
    \label{flow_contounous}
\end{figure}

We apply PFM to several tasks related to learning distributions of longitudinal data. One key starting point is the estimation of cross-sectional densities, which we refer to as panel densities, from irregularly observed data. This is useful, for example, for visualizing how the distribution of longitudinal measurements evolves over time.
We illustrate this task in Figure~\ref{fig: pan_illu}(A), where PFM estimates the panel densities from irregular samples by leveraging smoothness of the underlying densities over time.
This smoothness is incorporated into our continuous transport framework, yielding smooth density evolution across both longitudinal time and transport time, as illustrated in Figure~\ref{flow_contounous}. We show that such transport adapts to the true distribution of the observed data, enabling density estimation for general longitudinal data. Furthermore, we establish statistical convergence for panel density estimation under mild distributional conditions, revealing a phase-transition phenomenon that commonly arises in functional data analysis \citep{cai2011optimal,hsing2015theoretical,yan2025deep} and highlighting the benefits of PFM.

Building on the panel-flow representation, we further apply PFM to longitudinal data synthesis, completion, and classification, as illustrated in Figure~\ref{fig: pan_illu}(B)--(D). A key advantage of our approach is that it can estimate the joint data distribution and generate data on an arbitrary user-specified time grid based on irregular observations,  avoiding the preliminary dimension-reduction step (e.g., FPCA) and Gaussian assumptions commonly used in these tasks \citep{hsing2015theoretical,kraus2015components,wang2016functional,xue2024optimal}.
In addition, the generative nature of PFM provides a synthetic viewpoint for longitudinal classification: it can generate a time-grid-adaptive dataset tailored to the classification of a subject's longitudinal data; see Figure~\ref{fig: pan_illu}(D). This viewpoint can help mitigate class-imbalance issues that commonly arise in classification \citep{ma2026synthetic}.

We provide extensive simulation studies for panel density estimation, longitudinal data generation, and longitudinal classification, and compare our method with existing competing approaches. The results demonstrate the superiority of PFM for both Gaussian and non-Gaussian longitudinal data and highlight the benefit of avoiding dimension reduction and Gaussian assumptions in longitudinal settings. In addition, we apply PFM to a vaginal microbiome longitudinal dataset \citep{costello2022vaginal}, which consists of irregular vaginal swab samples from 188 pregnancies, with each pregnancy assigned a term or preterm label indicating pregnancy status. Our method achieves accurate classification of pregnancy status based on the longitudinal swab trajectories and also reveals clear time-varying distributional differences between the term and preterm groups through panel density estimation. These results demonstrate the practical utility of PFM for learning distributions of longitudinal data.

The rest of the article is organized as follows. In Section~\ref{sec:method}, we introduce the panel flow matching framework for modeling longitudinal data. In Section~\ref{sec:est}, we present the forward flow-matching and backward kernel-fitting procedures for irregular and sparse longitudinal data. In Section~\ref{sec:application}, we discuss the resulting applications and statistical properties. In Section~\ref{sec:sim}, we conduct extensive simulations comparing PFM with existing density-estimation, generative, and classification methods. Section~\ref{sec:real} applies PFM to classification and panel density estimation for a vaginal microbiome longitudinal dataset. We conclude with a discussion in Section~\ref{sec:dis}.

\section{Methodology}\label{sec:method}

Let $\mathbb{I}(\cdot)$ denote the indicator function and $\bm{I}_p$ denote the $p\times p$ identity matrix. When there is no ambiguity, we abbreviate a functional object $f(\cdot)$ by $f$. Write $\bm{v}(\bm{z})=(v_1(\bm{z}),\ldots,v_p(\bm{z}))^\top\in\mathbb{R}^p$ for $\bm{z}=(z_1,\ldots,z_p)^\top\in\mathbb{R}^p$. The Jacobian matrix and divergence of $\bm{v}(\bm{z})$ are defined by
\(
\nabla_{\bm{z}}\bm{v}(\bm{z})
:=
\left(\frac{\partial v_k(\bm{z})}{\partial z_\ell}\right)_{1\le k,\ell\le p},
\ \text{and}\
\operatorname{div}_{\bm{z}}\bm{v}(\bm{z})
:=
\sum_{k=1}^p \frac{\partial}{\partial z_k}\, v_k(\bm{z}),
\)
respectively.
Let $\det(\cdot)$ denote the determinant of a square matrix and let \( \delta(\cdot\,;\cdot) \) denote the Dirac measure. The support of a continuous random variable \( X \) or a density $f$ is denoted by \( \supp(X) \) or $\supp(f)$. 
Let $\|\cdot\|$ and $\|\cdot\|_{2}$ denote the Euclidean norm of vectors and the $L^{2}$ norm of functions, respectively.
We use ``\( \overset{d}{=} \)'' to denote equality in distribution and use \( \mathcal{N}(a, b) \) to denote the normal distribution with mean \( a \) and variance \( b \). 
Let $\bm{A}\succeq \bm{B}$ denote the positive semidefinite order for symmetric matrices, i.e.,
$\bm{A}-\bm{B}$ is positive semidefinite. 
Denote $\|\cdot\|_{\infty}$ as the maximum absolute value of the entries of a vector\(/\)matrix and \(
\|\bm g\|_{L^\infty}:=\operatorname*{sup}_{\bm{x}}\|\bm g(\bm{x})\|_{\infty}
\).

Let $\bm{X}(t)$ be a continuous $p$-dimensional random vector indexed by time $t\in\mathcal{T}$, and let $\{p_t:\,t\in\mathcal{T}\}$ denote the densities of $\bm{X}(t)$, referred to as the panel densities.
In this section, we model $\bm{X}(\cdot)$ and $p_t$ under the latent representation
\begin{equation}\label{latent_model}
\bm{X}(t)=\bm{\phi}_t\big(\bm{Z}(t)\big),\ t\in \mathcal{T},
\end{equation}
where $\bm{Z}(t)\in\mathbb{R}^p$ is a latent process on $\mathcal{T}$ and
$\bm{\phi}_t:\mathbb{R}^p\to\mathbb{R}^p$ is a nonlinear map for each $t\in\mathcal{T}$. 
Here, $\bm{Z}(t)$ is assumed to follow a simple base distribution (e.g., Gaussian), and we allow the panel
densities $\{p_t:\,t\in\mathcal{T}\}$ to belong to a broad class of distributions, leading to flexible modeling. In the following, we propose a panel flow construction for $\bm{\phi}_t$ to achieve the model \eqref{latent_model}.

\subsection{Continuous Panel Flow}\label{sec:panel-flow}

We first introduce a panel-flow representation for the map $\bm{\phi}_t$ in
\eqref{latent_model}. For each $t\in\mathcal{T}$, we assume that $\bm{\phi}_t$ arises from a continuous normalizing flow \citep[CNF;][]{chen2018neural}. Specifically, there exists a family of maps $\{\bm{\phi}_{u,t}:\mathbb{R}^p\to\mathbb{R}^p:\ u\in[0,1]\}$ with $\bm{\phi}_{0,t}=\operatorname{id}$ such that
\begin{equation}\label{data_generation}
\bm{Z}_u(t)=\bm{\phi}_{u,t}\!\big(\bm{Z}_0(t)\big), \qquad t\in\mathcal{T},\ u\in[0,1],
\end{equation}
where $\bm{Z}_0(t)\sim p_{0,t}$ for a known base distribution $p_{0,t}$, and $\bm{Z}_1(t)\sim p_t$. We then define $\bm{\phi}_t:=\bm{\phi}_{1,t}$, so that the flow transports $p_{0,t}$ to $p_t$ as $u$ moves from $0$ to $1$.

The following definition constructs the maps $\bm{\phi}_{u,t}$ via an ordinary differential equation (ODE).
\begin{D}[Vector Field for Panel Flow]\label{def_flow}
Let $\bm{v}_{u,t}:\mathbb{R}^p\to\mathbb{R}^p$ be a $(u,t)$-dependent vector field.
The family of diffeomorphisms $\bm{\phi}_{u,t}:\mathbb{R}^p\to\mathbb{R}^p$ induced by $\bm{v}_{u,t}$ is defined for $u\in[0,1]$ by
\begin{equation}\label{flow_def}
\frac{\partial}{\partial u}\,\bm{\phi}_{u,t}(\bm{x})
=
\bm{v}_{u,t}\!\big(\bm{\phi}_{u,t}(\bm{x})\big),
\qquad
\bm{\phi}_{0,t}(\bm{x})=\bm{x}.
\end{equation}
For a functional sequence $\{\bm{Z}_u:\ u\in[0,1]\}$ with $\bm{Z}_u(t)\in\mathbb{R}^p$, let $p_{u,t}$ denote the density of $\bm{Z}_u(t)$.
We say that $\bm{v}_{u,t}$ generates the probability path $\{p_{u,t}:\ u\in[0,1],\,t\in\mathcal{T}\}$ if \eqref{data_generation} holds
for all $u\in[0,1]$ and $t\in\mathcal{T}$.
\end{D}
We refer to the above construction as a panel flow. Unlike object-level flows, which transport the entire function $\bm{Z}_0(\cdot)$ to $\bm{Z}_1(\cdot)$, our panel flow acts pointwise in time: for each $t\in\mathcal{T}$, $\bm{\phi}_{u,t}$ transports $\bm{Z}_0(t)$ to $\bm{Z}_1(t)$. As a result, the vector field acts on $\mathbb{R}^p$ rather than on a function space, substantially reducing model complexity. See Table~\ref{flow_comparison} for a comparison.

\begin{table}[h]
\centering
\caption{Comparison of flow structures under the setting of this article.}
\label{flow_comparison}
\renewcommand{\arraystretch}{1.15}
\setlength\tabcolsep{4pt}
\footnotesize
\begin{tabular}{l p{6.0cm} p{6.0cm}}
\hline
\textbf{Name} 
& \textbf{Object-level Flow \citep{liu2022flow,lipman2022flow,kerrigan2023functional}} 
& \textbf{Panel Flow (this work)} \\ 
\hline
\textbf{Vector field} 
& $\mathcal{V}_u : \mathcal{H}^p \to \mathcal{H}^p$, $u \in [0,1]$, where $\mathcal{H}^p$ is the product of $p$ Hilbert spaces $\mathcal{H}$ of functions on $\mathcal{T}$. 
& $\bm{v}_{u,t} : \mathbb{R}^p \to \mathbb{R}^p$, $u \in [0,1]$, $t \in \mathcal{T}$. \\
\hline
\end{tabular}
\end{table}

\paragraph*{Invertibility of Panel Flow.}
Another advantage of panel flow is that it yields an invertible map in \eqref{latent_model}, which enables recovery of the latent process. 
\begin{Theorem}[Invertibility of Panel Flow]\label{inverse_formula}
Assume that the vector field $\bm{v}_{u,t}(\bm{x})$ is uniformly bounded for $u\in[0,1]$,
$t\in\mathcal{T}$, and $\bm{x}\in\mathbb{R}^p$, and is continuous in $u\in[0,1]$. Moreover, suppose
that for any $\bm{x},\bm{y}\in\mathbb{R}^p$ and $t\in\mathcal{T}$,
\begin{equation}\label{lip_con}
\big\|\bm{v}_{u,t}(\bm{x})-\bm{v}_{u,t}(\bm{y})\big\|_2 \le L \,\|\bm{x}-\bm{y}\|_2,
\end{equation}
for some Lipschitz constant $L$ independent of $u$, $\bm{x}$, $\bm{y}$, and $t$. Then:
\begin{itemize}
\item[(a)]
For each $u\in[0,1]$ and $t\in\mathcal{T}$, the map $\bm{\phi}_{u,t}:\mathbb{R}^p\to\mathbb{R}^p$
in \eqref{flow_def} is a diffeomorphism. Moreover, the inverse of $\bm{\phi}_{1,t}$ satisfies
$\bm{\phi}_{1,t}^{-1}=\bm{\psi}_{1,t}$ for each $t\in\mathcal{T}$, where $\bm{\psi}_{u,t}$ is the
solution to
\begin{equation}\label{inver_flow_def}
\frac{\partial}{\partial u}\,\bm{\psi}_{u,t}(\bm{z})
=
-\,\bm{v}_{1-u,t}\!\big(\bm{\psi}_{u,t}(\bm{z})\big),
\qquad
\bm{\psi}_{0,t}(\bm{z})=\bm{z},
\quad \text{for almost every } \bm{z}\in\mathbb{R}^p.
\end{equation}
\item[(b)] If $\bm{Z}(\cdot)$ and $\bm{X}(\cdot)$ satisfy model~\eqref{latent_model} with
$\bm{Z}(\cdot)\overset{d}{=}\bm{Z}_0(\cdot)$, $\bm{X}(\cdot)\overset{d}{=}\bm{Z}_1(\cdot)$, and
$\bm{\phi}_t=\bm{\phi}_{1,t}$, where $\bm{Z}_0$, $\bm{Z}_1$, and $\bm{\phi}_{1,t}$ satisfy
\eqref{data_generation}, then
\(
\bm{\psi}_{1,\cdot}\big(\bm{X}(\cdot)\big)\ \overset{d}{=}\ \bm{Z}(\cdot).
\)
\end{itemize}
\end{Theorem}
Theorem~\ref{inverse_formula} shows that, under the Lipschitz condition, the inverse map $\bm{\psi}_{1,t}=\bm{\phi}_{1,t}^{-1}$ exists and can be computed from \eqref{inver_flow_def}. Hence, under \eqref{latent_model}, pulling back $\bm{X}(\cdot)$ through $\bm{\psi}_{1,\cdot}$ recovers the latent distribution of $\bm{Z}(\cdot)$.

\begin{Remark}[Comparison with Variational Autoencoders]\label{re_VAE}
A longitudinal variational autoencoder \citep[VAE;][]{ramchandran2021longitudinal} specifies a model through a latent process $\bm{Z}(\cdot)$:
\begin{equation*}
\bm{X}(t)\mid \bm{Z}(t), t \sim p_\theta(\cdot\mid \bm{Z}(t), t),
\end{equation*}
where $p_\theta$ is modeled by a decoder network, and an encoder $q_\theta(\bm{Z}(t)\mid \bm{X}(t), t)$ together with the decoder is used for variational inference. Here, the decoder can be viewed as a map from the base distribution to the target distribution, while the encoder is a map from the target distribution back to the base distribution.
While VAEs share similarities with model~\eqref{latent_model}, they typically treat the decoder and encoder as distinct objects and estimate them separately through a variational approximation. In contrast, our panel flow represents both the forward map $\bm{\phi}_{1,t}$ and the inverse map $\bm{\psi}_{1,t}$ through the same vector field $\bm{v}_{u,t}$. This avoids approximation bias caused by potential inconsistency between separately estimated encoder and decoder maps.
\end{Remark}

\subsection{Target-Adaptive Panel Flow}

We now construct a vector field $\bm{v}_{u,t}$ that generates the panel flow by using a conditional vector field built from the base density $p_{0,t}$ and the target density $p_t$.

Specifically, given \( \bm{X}(t) \) and \(\bm{Z}_0(t)\), let \( q_{u,t}(\,\cdot \mid \bm{X}(t),\bm{Z}_0(t)) \) denote a conditional density that satisfies
\begin{eqnarray}\label{flow_conditionn}
    q_{0,t}(\cdot\mid \bm{X}(t),\bm{Z}_0(t)) = \delta(\cdot; \bm{Z}_0(t))
    \quad \text{and} \quad
    q_{1,t}(\cdot\mid \bm X(t),\bm{Z}_0(t)) = \delta(\cdot; \bm{X}(t)),
    \quad t \in \mathcal{T}.
\end{eqnarray}
Given \( q_{u,t} \), \(p_t\), and \(p_{0,t}\), define the marginal probability path \(p_{u,t}\) by
\begin{eqnarray}\label{pro_path}
    p_{u,t}(\cdot) := \int_{\mathbb{R}^p}\int_{\mathbb{R}^p}
    q_{u,t}(\cdot\mid \bm{x},\bm{z})\, p_t(\bm{x})p_{0,t}(\bm{z})\,
    \mathrm{d}\bm x \mathrm{d}\bm z.
\end{eqnarray}
Then \(p_{{u=0,t}}=p_{0,t}\) and \(p_{{u=1,t}}=p_t\), so \(\{p_{u,t}:u\in[0,1],\,t\in\mathcal{T}\}\) defines a path from the base density \(p_{0,t}\) to the target density \(p_t\).

\begin{Theorem}[Conditional-to-Marginal Panel Flow]\label{Theo: cond}
Let \( \bm{v}_{u,t}(\bm{y} \mid \bm{x},\bm{z} ) \) be a vector field that generates the conditional probability path \( \{q_{u,t}(\cdot \mid \bm{x},\bm{z});\, u \in [0,1] \} \) for each \( t \in \mathcal{T} \). 
For \(\bm{y}\in\supp(p_{u,t})\), define
\begin{eqnarray}
\label{Mar_vectorfield}
 \bm{v}_{u,t}(\bm{y}) 
= \int_{\mathbb{R}^p}\int_{\mathbb{R}^p} 
\bm{v}_{u,t}(\bm{y} \mid \bm x,\bm{z}) \cdot 
\frac{q_{u,t}(\bm{y} \mid \bm{x},\bm{z}) \cdot p_t(\bm{x})p_{0,t}(\bm{z})}{p_{u,t}(\bm{y})}\, \mathrm{d}\bm x \mathrm{d}\bm z.
\end{eqnarray}
If \( \{\bm Z_u(t);\, u \in [0,1] \} \) can be generated by \( \bm{v}_{u,t} \) with \( \bm Z_0(t) \sim p_{0,t} \), then \(\bm Z_u(t) \sim p_{u,t} \) for all \( u \in [0,1] \) and \( t \in \mathcal{T} \). Likewise, if \( \{\bm Z_u(t);\, u \in [0,1] \} \) is generated by \( \bm{v}_{u,t} \) via \eqref{inver_flow_def}, with the initial value \( \bm Z_0(t) \sim p_{1,t} \), then \(\bm Z_u(t) \sim p_{1-u,t} \) for all \( u \in [0,1] \) and \( t \in \mathcal{T} \). 
\end{Theorem}
Theorem~\ref{Theo: cond} provides a general construction of a panel flow from the base density $p_{0,t}$ to the target density $p_t$. A convenient choice of the conditional density in \eqref{flow_conditionn} is the rectified flow
\[
q_{u,t}(\cdot\mid \bm{x},\bm{z})
=
\delta\!\left(\cdot;(1-u)\bm{z}+u\bm{x}\right),
\qquad u\in[0,1],\ t\in\mathcal{T},
\]
which transports $\bm{z}$ to $\bm{x}$ along the linear interpolation $(1-u)\bm{z}+u\bm{x}$. The corresponding conditional vector field is
\[
\bm{v}_{u,t}(\bm{y}\mid \bm{x},\bm{z})=\bm{x}-\bm{z},
\qquad u\in[0,1],\ t\in\mathcal{T}.
\]
Applying Theorem~\ref{Theo: cond} then yields the vector field $\bm{v}_{u,t}$ for the path $\{p_{u,t}:u\in[0,1]\}$.

\begin{Remark}[Comparison with Karhunen--Lo\`eve Expansion]\label{remark_flow_object}
A classical approach to model {longitudinal} data is the Karhunen--Lo\`eve (KL) expansion
\citep{hsing2015theoretical,wang2016functional}:
\begin{equation}\label{KL_expansion}
X_{l}(t)=\mu_l(t)+\sum_{k=1}^{K}\xi_{lk}\,{\psi}_{lk}(t),
\qquad t\in\mathcal{T},\ \ l=1,\ldots,p,
\end{equation}
where $(X_1(t),\ldots,X_p(t))^\top=\bm{X}(t)$, $\mu_l(t)$ is the mean function of the $l$th
component, $\{\psi_{lk}\}_{k\ge 1}$ are the eigenfunctions, and $\{\xi_{lk}\}_{k\ge 1}$ are the
corresponding scores. In practice, the expansion is truncated at $K$ to obtain a low-rank representation of functions,
which forms the basis of functional data analysis and many subsequent modeling and inference procedures
\citep{wang2016functional}.
However, the KL expansion is essentially a linear latent model, in which the mean and eigenfunctions act as time-varying linear maps for the latent variables; consequently, the induced panel densities $\{p_t:t\in\mathcal{T}\}$ are constrained by this linear structure. In addition, common KL-based implementations often rely on Gaussian assumptions {for} $\bm{X}(\cdot)$ \citep{yao2005functional,xiao2018fast}, which may fail to capture complex features such as multimodality, {heavy tails}, and other general distributional characteristics.
In contrast, our panel flow introduces the nonlinear latent model~\eqref{latent_model}, yielding a target-adaptive flow map $\bm{\phi}_t$ and thereby providing greater flexibility for longitudinal modeling.
\end{Remark}

\subsection{Distribution Representation via Panel Flow}

We now describe how panel flow yields a representation of the distribution of $\{\bm{X}(t):t\in\mathcal{T}\}$. Assume that the marginal base density $p_{0,t}$ is known and that we observe samples of $\bm{X}(T)$ at random times $T\in\mathcal{T}$. The key steps are to estimate the vector field $\bm{v}_{u,t}$ from $\bm{X}(T)$ and then pull the observations back to the latent space to infer the joint distribution of $\{\bm{Z}(t):t\in\mathcal{T}\}$.

To obtain the vector field, we construct the flow-matching loss based on \eqref{flow_def}:
\begin{equation*}
\mathcal{L}(\bm{w})
:=
\mathbb{E}\bigg\|
\frac{\partial}{\partial u}\,\bm{\phi}_{u,T}\!\big(\bm{Z}_0(T)\big)\Big|_{u=U}
-
\bm{w}_{U,T}\!\Big(\bm{\phi}_{{U},T}\!\big(\bm{Z}_0(T)\big)\Big)
\bigg\|_2^2,
\end{equation*}
where $\bm{Z}_0(t)\sim p_{0,t}$ for $t\in\mathcal{T}$, $U\sim \text{Unif}([0,1])$, and $\{\bm{\phi}_{u,t}:u\in[0,1]\}$ is a flow
that transports the base density $p_{0,t}$ to the target panel density $p_t$ of $\bm{X}(t)$, for each
$t\in\mathcal{T}$. 
To compute the loss, we first generate samples of $\bm{\phi}_{U,T}\!\big(\bm{Z}_0(T)\big)$ conditional on
$\bm{X}$, $\bm{Z}_0$, $U$, and $T$ via the conditional density
$q_{u,t}(\,\cdot \mid \bm{X}(t),\bm{Z}_0(t))$, i.e., $\bm{\phi}_{U,T}\!\big(\bm{Z}_0(T)\big)\mid \bm{X},\bm{Z}_0,T
\sim q_{U,T}\!\left(\,\cdot \mid \bm{X}(T),\bm{Z}_0(T)\right).$
By Theorem~\ref{Theo: cond}, there exists a vector field $\bm{v}_{u,t}$ that transports $p_{0,t}$ to $p_t$ for each $t\in\mathcal{T}$. This vector field can be
learned by minimizing the loss $\mathcal{L}(\bm{w})$ over $\bm{w}$ by plugging in the sampled
$\bm{\phi}_{U,T}\!\big(\bm{Z}_0(T)\big)$, as formulated below.

\begin{Theorem}[Panel Flow Matching for Density Estimation]\label{thm:cnf_logdensity}
Let
$\bm{Z}_U(T)\mid \bm{X}(T),\bm{Z}_0(T),T,U \sim q_{U,T}\!\left(\,\cdot \mid \bm{X}(T),\bm{Z}_0(T)\right)$. Define
\begin{equation}\label{FM_loss}
\bm{v}
:=
\argmin_{\bm{w}}\mathcal{L}(\bm{w})
=
\argmin_{\bm{w}}
\mathbb{E}\bigg\|
\frac{\partial \bm{Z}_u(T)}{\partial u}\Big|_{u=U}
-
\bm{w}_{U,T}\!\Big(\bm{Z}_U(T)\Big)
\bigg\|_2^2.
\end{equation}
For any fixed $t\in\mathcal{T}$ and
$\bm{x}\in\mathbb{R}^p$, the target density satisfies
\begin{equation}\label{eq:cnf_logdensity}
\log p_t(\bm{x})
=
\log p_{0,t}(\bm{z}_0)
-
\int_0^1
\operatorname{div}_{\bm{x}}\bm{v}_{u,t}(\bm{z}_u)\,\mathrm{d}u,
\end{equation}
where \(\{\bm z_u:u\in[0,1]\}\) is a path ending at \(\bm x\), defined by
\(
\bm z_u=\bm\psi_{1-u,t}(\bm x).
\)
Here, the map \(\bm\psi_{u,t}\) is obtained by solving the inverse-flow equation~\eqref{inver_flow_def} with initial value \(\bm x\).
\end{Theorem}

Equation~\eqref{eq:cnf_logdensity} shows that the panel density $p_t(\bm{x})$ can be transported from the base density $p_{0,t}(\bm{z}_0)$ via an integral equation, where $\bm{z}_0$ is the pullback value associated with $\bm{x}$ via the vector field $\bm{v}_{u,t}$. An illustration of the density transport is given in Figure~\ref{flow_contounous}.

We next estimate the joint density of the base {$\big(\bm{Z}(t_1)^\top,\ldots,\bm{Z}(t_J)^\top\big)^\top$}, denoted by $p_{\bm{Z},t_1,\ldots,t_J}$, for given time points $t_1,\ldots,t_J\in\mathcal{T}$. To do so, we obtain pullback samples
\begin{eqnarray}\label{pull_back_sample}
  \bm{Z}(t):=  \bm{\psi}_{1,t}\big(\bm{X}(t)\big),\ t\in \mathcal{T},
\end{eqnarray}
and then estimate the joint density of $\bm{Z}(\cdot)$ based on these latent samples, as justified by Theorem~\ref{inverse_formula}(b).

\begin{Remark}[Full Density Representation]
A direct application of the above procedure is to represent the joint distribution of $\{\bm{X}(t);t\in \mathcal{T}\}$ based on $\bm{v}_{u,t}$ and  $p_{\bm{Z},t_1,\ldots,t_J}$. 
For any time grid $t_1,\ldots,t_J\in\mathcal{T}$ and under the representation $\bm{X}(t)=\bm{\phi}_{1,t}\big(\bm{Z}(t)\big)$, the joint density of
$\big(\bm{X}(t_1)^\top,\ldots,\bm{X}(t_J)^\top\big)^\top$
admits a change-of-variables form.
Define
the stacked vector
$\bm{x}_{1:J}:=\big(\bm{x}(t_1)^\top,\ldots,\bm{x}(t_J)^\top\big)^\top\in\mathbb{R}^{pJ}$.
Then the full density of $\big(\bm{X}(t_1)^\top,\ldots,\bm{X}(t_J)^\top\big)^\top$ is computed as
\begin{equation}\label{joint_dis}
p_{\bm{X},t_1,\ldots,t_J}(\bm{x}_{1:J})
=
p_{\bm{Z},t_1,\ldots,t_J}\!\Big(\bm{\psi}_{1,t_1}\{\bm{x}(t_1)\},\ldots,\bm{\psi}_{1,t_J}\{\bm{x}(t_J)\}\Big)
\prod_{j=1}^J
\left|\det \nabla_{\bm{x}}\bm{\psi}_{1,t_j}\!\big(\bm{x}(t_j)\big)\right|,
\end{equation}
where the inverse map $\bm{\psi}_{1,t}(\bm{x})$ and the associated Jacobian determinant
$\det \nabla_{\bm{x}}\bm{\psi}_{1,t}\!\big(\bm{x}\big)$ can be obtained via the pullback
equation~\eqref{inver_flow_def} with initial value $\bm{x}$. Specifically, for a fixed $t\in\mathcal{T}$
and $\bm{x}\in\mathbb{R}^p$, 
the log-Jacobian is computed as
\begin{equation*}
\log\left|\det \nabla_{\bm{x}}\bm{\psi}_{1,t}\!\big(\bm{x}\big)\right|
=
-\int_0^1 \operatorname{div}_{\bm{x}}\bm{v}_{1-u,t}\!\big(\bm{\psi}_{u,t}(\bm{x})\big)\,\mathrm{d}u.
\end{equation*}
See Part \ref{prf: theorem 3} of Supplementary Materials for the proof.
\end{Remark}

\section{Learning Panel Flow via Matching}\label{sec:est}

Consider a sample of $n$ independent and identically distributed subjects $\bm{X}_i$, $i=1,\ldots,n$. 
For each subject $i=1,\ldots,n$, let $\bm{Y}_{ij}\in\mathbb{R}^p$ be a measurement of
$\bm{X}_i(T_{ij})$, representing the observed values of $p$ features for subject $i$ at time
$T_{ij}\in\mathcal{T}$, where $j=1,\ldots,J_i$. The observation times
$\{T_{ij}:j=1,\ldots,J_i\}$ may vary across subjects, allowing for irregular sampling over the domain $\mathcal{T}$.

In this section, we implement the estimation of panel flows for irregular data
$\{(T_{ij},\bm{Y}_{ij}): i=1,\ldots,n,\ j=1,\ldots,J_i\}$ using neural networks.
For notation, let $\mathcal{F}_{d_{\mathrm{in}},d_{\mathrm{out}}}(L,m)$ denote the class of
fully connected neural networks $f:\mathbb{R}^{d_{\mathrm{in}}}\to\mathbb{R}^{d_{\mathrm{out}}}$
with $L$ hidden layers and width $m$. Concretely, any
$f\in\mathcal{F}_{d_{\mathrm{in}},d_{\mathrm{out}}}(L,m)$ admits the representation $f(\bm{s})
=
\bm{W}_{L+1}\,\sigma\!\Big(\bm{W}_{L}\,\sigma\big(\cdots \sigma(\bm{W}_1\bm{s}+\bm{b}_1)\cdots\big)
+\bm{b}_{L}\Big)
+\bm{b}_{L+1}$,
where $\bm{W}_1\in\mathbb{R}^{m\times d_{\mathrm{in}}}$,
$\bm{W}_{L+1}\in\mathbb{R}^{d_{\mathrm{out}}\times m}$,
$\bm{W}_{\ell}\in\mathbb{R}^{m\times m}$ for $\ell=2,\ldots,L$, $\bm{b}_\ell$ are bias vectors of
compatible dimensions, and $\sigma(\cdot)$ is an elementwise activation function, set as the ReLU
function $\sigma(x)=\max\{0,x\}$ in this article.

\subsection{Forward Flow Matching}\label{sec:ffm}

We construct an empirical flow matching loss based on \eqref{FM_loss} using the rectified
flow, i.e., $q_{u,t}(\cdot\mid \bm{x},\bm{z})=\delta\!\left(\cdot;(1-u)\bm{z}+u\bm{x}\right)$. As a result, $\bm{Z}_u(T_{ij})$ in \eqref{FM_loss}
is $(1-u)\bm{Z}_0(T_{ij})+u\bm{Y}_{ij}$, and the associated loss becomes 
\begin{eqnarray}\label{exp_loss}
    \mathcal{L}(\bm{w})
:=
\mathbb{E}\bigg\|
\bm{Y}_{ij}-\bm{Z}_{0}(T_{ij})
-
\bm{w}\!\Big(U,T_{ij},(1-U)\bm{Z}_0(T_{ij})+U\bm{Y}_{ij}\Big)
\bigg\|_2^2,
\end{eqnarray}
where $U\sim \mathrm{Unif}([0,1])$ and $\bm{Z}_0(t)\sim p_{0,t}$ for $t\in\mathcal{T}$.
To optimize the loss, we apply stochastic optimization by 
approximating the expectation using Monte Carlo. Specifically, we sample a mini-batch of index pairs
$\{(i_b,j_b)\}_{b=1}^B$, where $i_b$ is sampled uniformly from $\{1,\ldots,n\}$ and, conditional on
$i_b$, $j_b$ is sampled uniformly from $\{1,\ldots,J_{i_b}\}$, and $B$ is the batch size. For each $(i_b,j_b)$, we draw an
independent $U_b\sim\mathrm{Unif}([0,1])$ and a base sample
$\bm{Z}_{0,b}\sim p_{0,T_{i_b j_b}}$. We then form the bridge state
$\bm{Z}_{u,b}:=(1-U_b)\bm{Z}_{0,b}+U_b\bm{Y}_{i_b j_b}$ and the corresponding velocity
$\bm{V}_{b}:=\bm{Y}_{i_b j_b}-\bm{Z}_{0,b}$. The mini-batch loss is then constructed as 
\begin{equation}\label{FM_loss_emp}
\widehat{\mathcal{L}}_B(\bm{w})
=
\frac{1}{B}\sum_{b=1}^B
\big\|
\bm{V}_{b}
-
\bm{w}\!\big(U_b,T_{i_b j_b},\bm{Z}_{u,b}\big)
\big\|_2^2.
\end{equation}

We then parameterize $\bm{w}(u,t,\bm{x})=\bm{\omega}_\theta\big(u,t,\bm{x}\big)$ for $\bm{\omega}_\theta\in\mathcal{F}_{2+p,p}(L,m)$, and apply gradient-based optimization to iteratively update $\theta$.
In detail, at each iteration, we resample a mini-batch of index pairs $\{(i_b,j_b)\}_{b=1}^B$, draw
$\{U_b\}_{b=1}^B$ and $\{\bm{Z}_{0,b}\}_{b=1}^B$, and construct the corresponding mini-batch loss
$\widehat{\mathcal{L}}_B(\bm{\omega}_\theta)$ at the current iterate $\theta$.
We then update $\theta$ by one step of adaptive moment estimation \citep[Adam;][]{kingma2014adam}
using the gradient $\nabla_\theta \widehat{\mathcal{L}}_B(\bm{\omega}_\theta)$, which performs
stochastic gradient descent on the expected loss in \eqref{exp_loss} with adaptive learning rates.
Repeating this update over many mini-batches
yields the final estimator $\widehat{\theta}$ and
\(
\widehat{\bm{v}}_{u,t}(\bm{x})
=
\bm{\omega}_{\widehat{\theta}}\big(u,t,\bm{x}\big).
\)

\begin{Remark}[Comparison to Object-Level Flow]
Object-level flow matching \citep{liu2022flow,lipman2022flow,kerrigan2023functional} requires evaluating a loss on entire trajectories $\bm{X}(\cdot)$ and learning a vector field on a function space (see Table~\ref{flow_comparison}). This is difficult to implement under irregular observation, where only pairs $\{(T_{ij},\bm{Y}_{ij})\}$ are available. In contrast, the panel flow matching loss \eqref{FM_loss_emp} is constructed directly from the observed time-value pairs and targets a lower-dimensional vector field on $\mathbb{R}^p$. This makes it better suited to longitudinal data observed on subject-specific time grids.
\end{Remark}

\begin{algorithm}[ht!]
\caption{Panel Flow Matching}
\label{alg:pfm_full}
\begin{algorithmic}[1]
\Require Panel data $\{(T_{ij},\bm{Y}_{ij}) : i=1,\ldots,n,\ j=1,\ldots,J_i\}$; network depth $L$, width $m$; batch size $B$; training steps $H$; {rank} $r$.
\Ensure Estimated vector field $\widehat{\bm{v}}_{u,t}(\bm{x})$; estimated kernel $\widehat{\bm{C}}(s,t)$.

\State Parameterize the vector field by $\bm{\omega}_\theta\in\mathcal{F}_{2+p,\,p}(L,m)$.
\State Parameterize the kernel factor by $\bm{g}_\gamma\in\mathcal{F}_{1,\,pr}(L,m)$:
$\bm{A}_\gamma(t):=\mathrm{mat}\big(\bm{g}_\gamma(t)\big)\in\mathbb{R}^{p\times r}$,
and $\bm{C}_\gamma(s,t):=\bm{A}_\gamma(s)\bm{A}^\top_\gamma(t)$.

\vspace{2pt}
\Statex \textbf{Stage I: Forward flow matching}
\State Initialize $\theta$.
\For{$s=1,2,\ldots,H$}
    \State Sample $\{(i_b,j_b)\}_{b=1}^B$ with $i_b\sim \mathrm{Unif}\{1,\ldots,n\}$ and $j_b\mid i_b\sim \mathrm{Unif}\{1,\ldots,J_{i_b}\}$. 
    \State Draw $\{U_b\}_{b=1}^B$ i.i.d.\ from $\mathrm{Unif}([0,1])$ and $\{\bm{Z}_{0,b}\}_{b=1}^B$ i.i.d.\ from $\mathcal N(\bm{0},\bm{I}_p)$.
    \State Set $T_b:=T_{i_b j_b}$, $\bm{Z}_{u,b}:=(1-U_b)\bm{Z}_{0,b}+U_b\bm{Y}_{i_b j_b}$, and $\bm{V}_b:=\bm{Y}_{i_b j_b}-\bm{Z}_{0,b}$, $b=1,\ldots,B$.
    \State Compute mini-batch loss:
    \(
    \widehat{\mathcal{L}}_{B}(\theta)
    :=
    \frac{1}{B}\sum_{b=1}^B
    \big\|
    \bm{V}_b - {\bm{\omega}_\theta}(U_b,T_b,\bm{Z}_{u,b})
    \big\|_2^2.
    \)
    \State Update $\theta$ by one Adam step using $\nabla_\theta \widehat{\mathcal{L}}_{B}(\theta)$.
\EndFor
\State Set $\widehat{\bm{v}}_{u,t}(\bm{x}) := {\bm{\omega}_{\theta}}(u,t,\bm{x})$.

\Statex \textbf{Stage II: Backward kernel fitting}
\State Compute $\widehat{\bm{Z}}_{ij}:=\widehat{\bm{\psi}}_{1,T_{ij}}(\bm{Y}_{ij})$ by solving \eqref{inver_flow_def} with $\bm{v}_{u,t}=\widehat{\bm{v}}_{u,t}$ and initial value $\bm{Y}_{ij}$. 
\State Initialize $\gamma$.
\For{$s=1,2,\ldots,H$}
    \State Sample $\{i_b\}_{b=1}^B$ with $i_b\sim \mathrm{Unif}\{1,\ldots,n\}$.
    \State Sample $j_{b,1},j_{b,2}\mid i_b \sim \mathrm{Unif}\{1,\ldots,J_{i_b}\}$ independently for $b=1,\ldots,B$.
    \State Set $S_b:=T_{i_b j_{b,1}}$, $T_b:=T_{i_b j_{b,2}}$, and $\bm{M}_b:=\widehat{\bm{Z}}_{i_b j_{b,1}}\widehat{\bm{Z}}_{i_b j_{b,2}}^\top$ for $b=1,\ldots,B$.
    \State Compute mini-batch loss:
    \(
    \widehat{\mathcal{L}}^{\mathrm{cov}}_{B}(\gamma)
    :=
    \frac{1}{B}\sum_{b=1}^B
    \big\|
    \bm{M}_b - \bm{A}_\gamma(S_b)\bm{A}_\gamma(T_b)^\top
    \big\|_F^2.
    \)
    \State Update $\gamma$ by one Adam step using $\nabla_\gamma \widehat{\mathcal{L}}^{\mathrm{cov}}_{B}(\gamma)$.
\EndFor
\State Output $\widehat{\bm{C}}(s,t):=\bm{A}_{\gamma}(s)\bm{A}^\top_{\gamma}(t)$.

\end{algorithmic}
\end{algorithm}

\subsection{Backward Kernel Fitting}

Given the estimated vector field $\widehat{\bm{v}}_{u,t}$, we pull back each observed sample
$\bm{Y}_{ij}$ to the latent space by solving \eqref{inver_flow_def}, with $\bm{v}_{u,t}$ replaced by
$\widehat{\bm{v}}_{u,t}$. Denote the pullback samples by
\begin{equation*}
\widehat{\bm{Z}}_{ij}:=\widehat{\bm{\psi}}_{1,T_{ij}}(\bm{Y}_{ij}), \qquad i=1,\ldots,n,\ \ j=1,\ldots,J_i,
\end{equation*}
where $\widehat{\bm{\psi}}_{1,t}$ is the inverse map induced by $\widehat{\bm{v}}_{u,t}$ through
\eqref{inver_flow_def}. This subsection focuses on estimating the distribution of the latent variables under a Gaussian
assumption, where latent Gaussianity is a commonly used assumption in VAEs
\citep{ramchandran2021longitudinal} and in sparse functional data analysis
\citep{yao2005functional,hall2008modelling}.
To this end, we take $p_{0,t}$,
$t\in\mathcal{T}$, to be $\mathcal N(0,\bm{I}_p)$ for obtaining the vector field $\widehat{\bm{v}}_{u,t}$ and the resulting pullback samples $\widehat{\bm{Z}}_{ij}$. 

Under the Gaussian assumption, the latent distribution is fully determined by the {positive semidefinite} 
kernel $\bm{C}(s,t)=\mathbb{E}\{\bm{Z}(s)\bm{Z}^\top(t)\}\in\mathbb{R}^{p\times p}$, where $\bm{Z}$ is the latent process
{corresponding to the pullback samples in \eqref{pull_back_sample}}. Given this, we construct the loss
\begin{equation*}
\mathcal{L}_{\mathrm{cov}}(\bm{K})
:=
\mathbb{E}\Big\|
\bm{Z}(S)\bm{Z}^\top(T)
-
\bm{K}(S,T)
\Big\|_F^2,
\end{equation*}
where $S$ and $T$ are random times in $\mathcal{T}$ independent of $\bm{Z}$, and $\|\cdot\|_F$
denotes the Frobenius norm. By the definition of conditional expectation, the minimizer $\bm{K}^\star$
of $\mathcal{L}_{\mathrm{cov}}(\bm{K})$ over measurable $\bm{K}$ satisfies $\bm{K}^\star(s,t)
=
\mathbb{E}\big\{\bm{Z}(S)\bm{Z}^\top(T)\mid S=s,T=t\big\}$, which is exactly
the true kernel $\bm{C}(s,t)$. We therefore use the loss to infer the kernel $\bm{C}(s,t)$.

To parameterize the kernel $\bm{K}(s,t)$, we adopt the factorized form
\(
\bm{K}(s,t)= \bm{A}(s)\bm{A}^\top(t),\ \bm{A}(t)\in\mathbb{R}^{p\times r},
\)
where $r$ is a rank parameter of the kernel. This representation is motivated by two considerations. First, it
provides a convenient low-rank structure for modeling a time-varying covariance kernel, which reduces the number of parameters. Second, the factorization
automatically guarantees that $\bm{K}$ is positive semidefinite. We take $r\ge p$ so that the
contemporaneous covariance
\(
\bm{K}(t,t)=\bm{A}(t)\bm{A}^\top(t)
\)
can have full rank for all $t\in\mathcal{T}$.
Given this parameterization, we rewrite the loss as
\begin{equation*}
\mathcal{L}_{\mathrm{cov}}(\bm{A})
:=
\mathbb{E}\Big\|
\bm{Z}(S)\bm{Z}^\top(T)
-
\bm{A}(S)\bm{A}^\top(T)
\Big\|_F^2.
\end{equation*}
Similar to vector
field estimation, we approximate $\mathcal{L}_{\mathrm{cov}}(\bm{A})$ by
mini-batches 
$\big(\widehat{\bm{Z}}_{i_b j_{b,1}}\widehat{\bm{Z}}_{i_b j_{b,2}}^\top,\allowbreak\,
T_{i_b j_{b,1}},\,T_{i_b j_{b,2}}\big)$, $b=1,\ldots,B$, where $i_b$ is sampled uniformly from
$\{1,\ldots,n\}$ and, conditional on $i_b$, the indices $j_{b,1}$ and $j_{b,2}$ are sampled
uniformly from $\{1,\ldots,J_{i_b}\}$. Accordingly, we parameterize $\bm{A}(t)$ by a neural network, i.e.,
$\bm{A}(t):=\mathrm{mat}\big(\bm{g}_\gamma(t)\big)\in\mathbb{R}^{p\times r}$ with  $\bm{g}_{\gamma}\in\mathcal{F}_{1,\,pr}(L,m)$, where $\mathrm{mat}(\cdot)$ reshapes a length-$pr$ vector into a $p\times r$ matrix. Given this, we optimize the network parameters $\gamma$ using Adam based on the corresponding mini-batch
approximation of $\mathcal{L}_{\mathrm{cov}}(\bm{A})$. The complete panel flow matching procedure for vector field and kernel estimation is summarized in Algorithm~\ref{alg:pfm_full}.

Algorithm~\ref{alg:pfm_full} is applicable to noisy observations, i.e.,
$\bm{Y}_{ij}=(Y_{ij1},\ldots,Y_{ijp})^\top=\bm{X}_i(T_{ij})+\bm{\varepsilon}_{ij}$, where
$\bm{\varepsilon}_{ij}$ is {mean-zero} noise. To reduce the effect of noise, we prefer to implement a denoising step on $\bm{Y}_{ij}$ prior to PFM. In detail, if a subject has a moderate number of {observations} (e.g., $J_i\ge J_{\min}$ for some
threshold $J_{\min}$), we smooth each feature trajectory $\{Y_{ij\ell}:j=1,\ldots,J_i\}$ over
$\{T_{ij}\}_{j=1}^{J_i}$ using a smoothing spline \citep{gu2013smoothing}, and replace $Y_{ij\ell}$
by the fitted value at $T_{ij}$. This procedure reduces noise by exploiting the smoothness of the
underlying functions. 

In implementation, we set $J_{\min}=6$ in the denoising step and then apply Algorithm~\ref{alg:pfm_full} to the denoised panels, with $B=500$, $H=20000$, and $r=p$. The values of $L$ and $m$ for the neural networks may be chosen according to the total sample size $\sum_{i=1}^n J_i$, as suggested by Theorem~\ref{as:gao12_panel} below. For a moderate sample size, we recommend setting $L=3$ and $m=30$, as three-layer networks provide sufficient flexibility for fitting \citep{ismailov2023three}.

\section{Applications of PFM to Longitudinal Data}\label{sec:application}

This section applies panel flow matching to three longitudinal-data tasks: panel density estimation, trajectory synthesis/completion, and classification.

\subsection{Panel Density Estimation} 

Estimating panel densities $p_t$ provides a way to capture cross-sectional distributional patterns of longitudinal data, as illustrated in Figure~\ref{fig: pan_illu}. Panel density estimation can be carried out via \eqref{eq:cnf_logdensity} by replacing the vector field $\bm{v}_{u,t}$ with its estimate $\widehat{\bm{v}}_{u,t}$:
\begin{equation*}
 \log \hat p_t(\bm{x})
=
\log p_{0,t}\!\big(\hat{\bm{\psi}}_{1,t}(\bm{x})\big)
-
\int_0^{1}
\operatorname{div}_{\bm{x}}\hat{\bm{v}}_{u,t}\!\big(\hat{\bm{\psi}}_{1-u,t}(\bm{x})\big)\,\mathrm{d}u,
\end{equation*}
where $\hat{\bm{\psi}}_{u,t}(\bm{x})$ solves
\[
\frac{\partial}{\partial u}\,\hat{\bm{\psi}}_{u,t}(\bm{x})
=
-\,\hat{\bm{v}}_{1-u,t}\!\big(\hat{\bm{\psi}}_{u,t}(\bm{x})\big),
\qquad
\hat{\bm{\psi}}_{0,t}(\bm{x})=\bm{x}.
\]
Note that panel density estimation only requires vector-field estimation and does not require kernel estimation. Therefore, the latent Gaussian assumption is not needed for panel density estimation.

In the following, we study the statistical convergence of panel density estimation. We assume that the dimension $p$ is fixed, and, to simplify notation, suppose $J_i=J$ for all $i=1,\ldots,n$. We focus on the noiseless case where $\bm{Y}_{ij}=\bm{X}_i(T_{ij})$, $i = 1,\ldots,n$, $j = 1,\ldots,J$. Without loss of generality, $\mathcal{T}=[0,1]$. Suppose that \( \{T_{ij};\ i = 1,\ldots,n,\ j = 1, \ldots, J\} \) and \( \{\bm X_i;\ i = 1,\ldots,n\} \) are independent. For two sequences of non-negative real numbers $\{a_{nJ}\}$ and $\{b_{nJ}\}$, we write
$a_{nJ}\lesssim b_{nJ}$ (or equivalently $b_{nJ}\gtrsim a_{nJ}$) if there exists a constant $C>0$
independent of $(n,J)$ such that
\(
a_{nJ}\le C\,b_{nJ}
\ \text{for all }n,J.
\)
We write $a_{nJ}\asymp b_{nJ}$ if $a_{nJ}\lesssim b_{nJ}$ and $a_{nJ}\gtrsim b_{nJ}$.

To study the consistency of panel flow, we focus on the full-batch empirical objective induced by the rectified flow; see \eqref{FM_loss_emp}. For each observed pair
$(T_{ij},\bm{Y}_{ij})$, draw an independent $U_{ij}\sim\mathrm{Unif}([0,1])$ and an independent base
sample $\bm{Z}_{0,ij}$ from the standard Gaussian distribution, and define the bridge state and target velocity by
\(
\bm{Z}_{ij,U_{ij},T_{ij}}:=(1-U_{ij})\bm{Z}_{0,ij}+U_{ij}\bm{Y}_{ij},
\
\bm{V}_{ij}:=\bm{Y}_{ij}-\bm{Z}_{0,ij}.
\)
We consider the full-batch empirical risk minimization
\begin{equation}\label{FM_loss_full}
\widehat{\bm{v}}
=
\argmin_{\bm{w}\in\mathcal{W}_{L,m}}
\frac{1}{nJ}\sum_{i=1}^n\sum_{j=1}^J
\big\|
\bm{V}_{ij}
-
\bm{w}\!\big(U_{ij},T_{ij},\bm{Z}_{ij,U_{ij},T_{ij}}\big)
\big\|_2^2,
\end{equation}
where
\(
\mathcal{W}_{L,m}
:=
\Big\{\bm{f}\big(u,t,\bm{x}\big)
: \bm{f}\in \mathcal{F}_{2+p,p}(L,m),\ \|\bm{f}\|_{L^{\infty}}\leq C_1,\ \text{and}\ \allowbreak  \sup_{\bm{x}\neq \bm y}\big\|\bm{f}(u,t,\bm{x})-\bm{f}(u,t,\bm{y})\big\|_2 / \|\bm{x}-\bm{y}\|_2 \leq C_2
\Big\},
\)
with constants $C_1$ and $C_2$, and the resulting estimator is denoted by $\widehat{\bm{v}}_{u,t}(\bm{x})=\widehat{\bm{v}}(u,t,\bm{x})$.
Here, the boundedness conditions in $\mathcal{W}_{L,m}$ are imposed to guarantee that the flow map induced by $\widehat{\bm{v}}_{u,t}(\bm{x})$ is a diffeomorphism, as indicated by Theorem~\ref{inverse_formula}.

We impose the following assumptions for vector field estimation.

\begin{asum}\label{as:gao12_panel}
The panel densities $p_t$, $t\in\mathcal{T}$, are positive and uniformly bounded. Let $\bar{p}_t$ be the density of $\bm{X}(t)-\bm{\mu}(t)$, where $\bm{\mu}(t)=\mathbb{E}\bm{X}(t)$ and $\sup_{t\in \mathcal{T}}\|\bm{\mu}(t)\|_{\infty}<\infty$. Moreover, $\bar p_t$ satisfies one of the following conditions: (i) Strong log-concavity: There exist constants $\alpha,\beta>0$ such that for each
$t\in\mathcal{T}$, one can write
\(
\bar p_t(\bm{x}) \propto \exp\{-U_t(\bm{x})\}, \ \bm{x}\in\mathbb{R}^p,
\)
where $U_t:\mathbb{R}^p\to\mathbb{R}$ is twice continuously differentiable and satisfies
\(
\beta\,\bm{I}_p \ \succeq\ \nabla_{\bm{x}}^2 U_t(\bm{x}) \ \succeq\ \alpha\,\bm{I}_p,
\ \forall\,\bm{x}\in\mathbb{R}^p,\ \forall\,t\in\mathcal{T}.
\) 
(ii) bounded-support density: $\beta>0$ such that for each
$t\in\mathcal{T}$, one can write
\(
\bar p_t(\bm{x}) \propto \exp\{-U_t(\bm{x})\}, \ \bm{x}\in\mathbb{R}^p,
\)
where $U_t:\mathbb{R}^p\to\mathbb{R}$ is twice continuously differentiable and satisfies
\(
\beta\,\bm{I}_p \ \succeq\ \nabla_{\bm{x}}^2 U_t(\bm{x}) \ \succeq\ -\beta\,\bm{I}_p,
\ \forall\,\bm{x}\in\mathbb{R}^p,\ \forall\,t\in\mathcal{T}.
\) 
(ii) Mixtures of Gaussians: There exist constants $\sigma>0$ and $R< \infty$ such that for each $t\in\mathcal{T}$,
\(
\bar p_t=\mathcal{N}_{\sigma^2}\ast \rho_t,
\)
where $*$ denotes convolution of densities, $\mathcal{N}_{\sigma^2}$ is the density of $\mathcal N(\bm{0},\sigma^2\bm{I}_p)$, and $\rho_t$ is a
probability density supported on a Euclidean ball of radius at most $R$.
\end{asum}

\begin{asum}\label{asum:timepoint}
The time points \( \{T_{ij} \,;\, i = 1, \ldots, n,\; j = 1, \ldots, J\} \) are independently drawn from a density that is continuous and positive on \( \mathcal{T} \).
\end{asum}

\begin{asum}\label{as:t_smooth_pt}
The panel density $p_t(\bm{x})$ is jointly continuous in $(t,\bm{x})$ and differentiable in $t$. Moreover,
\(
\sup_{t\in \mathcal{T}}\int_{\mathbb{R}^p}\big|\partial_t p_t(\bm{x})\big|\,\mathrm{d}\bm{x}<\infty,
\ 
\sup_{t\in \mathcal{T}}\int_{\mathbb{R}^p}\|\bm{x}\|_2\big|\partial_t p_t(\bm{x})\big|\,\mathrm{d}\bm{x}<\infty.
\)
\end{asum}

Assumption~\ref{as:gao12_panel} imposes mild regularity on the panel densities of the centered process
$\bm{X}_i(t)-\bm{\mu}(t)$, including a broad class of well-behaved distributions
commonly assumed in flow matching \citep{gao2024convergence}. Assumption~\ref{asum:timepoint}
controls the sampling of observation times, which is standard in longitudinal/functional data analysis \citep{yao2005functional,hsing2015theoretical}. Assumption~\ref{as:t_smooth_pt}
is a basic requirement for PFM: it postulates sufficient time-smoothness of the panel
densities $\{p_t:t\in \mathcal{T}\}$ so that information can be borrowed across time points when observations are irregular.

Denote the true vector field $\bm{v}_{u,t}(\bm{x})$ as the minimizer of the loss
\eqref{exp_loss}, with the base density $p_{0,t}$ set as the standard Gaussian
density for $t\in\mathcal{T}$. We have the following convergence result.

\begin{Theorem}[Consistency of Vector Field Estimation]\label{thm:pfm_rate_like31}
Suppose Assumptions~\ref{as:gao12_panel}--\ref{as:t_smooth_pt} hold. Set
\(
m \ \asymp\ I_1\log I_1,
\
L \ \asymp\ I_2\log I_2,
\)
where $I_1$ and $I_2$ satisfy
\(
I_1I_2
\ \asymp\
(nJ)^{(2+p)/(8+2p)}\,
\big(\log(nJ)\big)^{-(8+4p)/(4+p)}.
\)
Recall that $p_{u,t}$ is the density of $\bm{Z}_{ij,u,t}$.
Then, for any fixed $A>0$ and $\tau\in(0,1)$, 
$\bm{v}_{u,t}$ is a globally Lipschitz continuous vector field with respect to $(u,t,\bm{x})$ on $[0,1-\tau]\times\mathcal{T}\times\mathbb{R}^p$, and
\begin{eqnarray}\label{rate_vector}
&&\mathbb{E}\bigg[
\int_{0}^{1-\tau}\int_{\mathcal{T}}\int_{[-A,A]^p}
\big\|\widehat{\bm{v}}_{u,t}(\bm{x})-\bm{v}_{u,t}(\bm{x})\big\|_{\infty}^2\,
p_{u,t}(\bm{x})\ \mathrm{d}\bm{x}\,\mathrm{d}t\,\mathrm{d}u
\bigg]\nonumber\\
&\lesssim&
n^{-1}+(nJ)^{-2/(4+p)}\big(\log(nJ)\big)^{16/(4+p)}.
\end{eqnarray}
As a result, $\mathbb{E}\bigg[
\int_{0}^{1}\int_{\mathcal{T}}\int_{\mathbb{R}^p}
\big\|\widehat{\bm{v}}_{u,t}(\bm{x})-\bm{v}_{u,t}(\bm{x})\big\|_{\infty}^2\,
p_{u,t}(\bm{x})\ \mathrm{d}\bm{x}\,\mathrm{d}t\,\mathrm{d}u
\bigg]=o(1)$, as $n\rightarrow\infty$. 
\end{Theorem}

In Theorem~\ref{thm:pfm_rate_like31}, the truncations $A$ and $\tau$ account for (i) the unboundedness of the transport samples
$\bm{Z}_{ij,U_{ij},T_{ij}}$ and (ii) the exploding Lipschitz constant in the $u$-direction
for $\bm{v}_{u,t}(\bm{x})$ as $u\uparrow 1$, as indicated by
\citet{gao2024convergence}. The resulting truncation error is asymptotically negligible, which yields consistency of the vector-field estimator and, in turn, supports consistent panel density estimation.

The error bound \eqref{rate_vector} provides a neural-network regression result for estimating the panel-flow vector field. It shows that the error bound exhibits a phase transition phenomenon that commonly arises in functional data analysis \citep{hsing2015theoretical,yan2025deep}. Ignoring logarithmic factors, the phase transition occurs at the scaling
\(
n^{-1}\ \asymp\ (nJ)^{-2/(4+p)}
\quad\Longleftrightarrow\quad
J\ \asymp\ n^{(2+p)/2}.
\)
Consequently, when $J \lesssim n^{(2+p)/2}$, the rate is essentially $(nJ)^{-2/(4+p)}$; otherwise, the rate is dominated by $n^{-1}$.
This transition indicates that when the time grid is sufficiently dense, the rate reduces to the parametric rate in $n$, whereas in the sparse regime, the error is governed by the nonparametric rate in $nJ$.

This phase transition is caused by the within-subject time dependence of $\bm{Y}_{ij}$ across different $j$.
When such dependence disappears, that is, when $\bm{Y}_{ij}$ are independent across both $i$ and $j$, the ``subject-level'' term $n^{-1}$ in \eqref{rate_vector} disappears, and the rate improves to $(nJ)^{-2/(4+p)}$ \citep{gao2024convergence}.
In this case, the error for vector field estimation depends primarily on the product $nJ$, and less on $n$ and $J$ separately.

Given the consistency of the vector field, we establish the corresponding result for panel density estimation.

\begin{Theorem}[Consistency of Panel Density Estimation]\label{thm:pfm_rate_like32}
Suppose the conditions in Theorem~\ref{thm:pfm_rate_like31} hold. Assume
\begin{eqnarray}\label{densiyu_lip}
    \sup_{t\in\mathcal{T}}\int_{\mathbb{R}^p}\|\nabla_{\bm{x}}\log p_t(\bm{x})\|_2^2\,p_t(\bm{x})\,\mathrm{d}\bm{x}
<\infty.
\end{eqnarray}
Define
\(
\mathcal{R}_{n,J}
:=
\mathbb{E}\bigg[
\int_{0}^{1}\int_{\mathcal{T}}\int_{\mathbb{R}^p}
\big\|\widehat{\bm{v}}_{u,t}(\bm{x})-\bm{v}_{u,t}(\bm{x})\big\|_{\infty}^2\,
p_{u,t}(\bm{x})\ \mathrm{d}\bm{x}\,\mathrm{d}t\,\mathrm{d}u
\bigg].
\)
As $n\to \infty$,
\[
\mathbb{E}\int_{\mathcal{T}}\int_{\mathbb{R}^p}
\Big|\log \hat p_t(\bm{x})-\log p_t(\bm{x})\Big|\,p_t(\bm{x})\,\mathrm{d}\bm{x}\,\mathrm{d}t
=
O\!\big(\mathcal{R}_{n,J}^{1/2}\big).
\]
\end{Theorem}

Condition~\eqref{densiyu_lip} controls the Lipschitz continuity of the log density $p_t(\bm{x})$ with
respect to $\bm{x}$ across time $t$. Under this, the density estimation error is controlled by
the vector-field estimation error, with rate
\(
n^{-1/2}+(nJ)^{-1/(p+4)},
\)
as implied by Theorem~\ref{thm:pfm_rate_like31}.

Generally, estimating a Lipschitz continuous density from vector data achieves the nonparametric rate
$n^{-1/(p+2)}$ \citep{takezawa2005introduction,jiang2017uniform}, where $n$ is the sample size and
$p$ is the data dimension. Compared to this, the rate
\(
n^{-1/2}+(nJ)^{-1/(p+4)}
\)
exhibits additional complexity due to within-subject dependence and the smooth time-varying structure
of the density. Meanwhile, this structure improves the effective sample size in the
nonparametric term from $n$ to $nJ$, highlighting the benefit of PFM for pooling temporal information
in panel density estimation.

\subsection{Data Synthesis and Completion}
Algorithm~\ref{alg:pfm_full} is also applicable to longitudinal data completion and to generating new longitudinal trajectories over a given time support.
In detail, fix a time grid $t_1,\ldots,t_J\in\mathcal{T}$. We first sample latent
panels
\(
\bm{z}_{1:J}^{(f)}:=\big(\bm{z}^{(f)}(t_1)^\top,\ldots,\bm{z}^{(f)}(t_J)^\top\big)^\top, \ f=1,\ldots,F,
\)
from a latent distribution, where $F$ is the number of generated trajectories. Then, for each
$f=1,\ldots,F$ and each $j=1,\ldots,J$, we generate $\tilde{\bm{X}}^{(f)}(t_j)$ by pushing
$\bm{z}^{(f)}(t_j)$ forward through the learned vector field $\widehat{\bm{v}}_{u,t}$ from Algorithm~\ref{alg:pfm_full}: solve
\(
\frac{\mathrm{d}}{\mathrm{d}u}\bm{x}^{(f)}_{u}(t_j)
=
\widehat{\bm{v}}_{u,t_j}\!\big(\bm{x}^{(f)}_{u}(t_j)\big)
\)
with $\bm{x}^{(f)}_{0}(t_j)=\bm{z}^{(f)}(t_j)$,
and set $\tilde{\bm{X}}^{(f)}(t_j):=\bm{x}^{(f)}_{1}(t_j)$.
The collections
$\{\tilde{\bm{X}}^{(f)}(t_j): j=1,\ldots,J\}$, $f=1,\ldots,F$, form generated trajectories on the
grid.

\begin{algorithm}[ht!]
\caption{Synthesis and Completion of Longitudinal Data via Panel Flow Matching}
\label{alg:pfm_synthetic}
\begin{algorithmic}[1]
\Require Panel data $\{(T_{ij},\bm{Y}_{ij}) : i=1,\ldots,n,\ j=1,\ldots,J_i\}$; grid $t_1,\ldots,t_J\in\mathcal{T}$; sample size $F$; optional subject index $i^\star$.
\Ensure Synthetic trajectories $\{\tilde{\bm{X}}^{(f)}(t_j): j=1,\ldots,J\}_{f=1}^F$; if $i^\star$ is given, completion samples $\{\tilde{\bm{X}}^{(f)}_{i^\star}(t_j): j=1,\ldots,J\}_{f=1}^F$.

\State Apply Algorithm~\ref{alg:pfm_full} to estimate $\widehat{\bm{v}}_{u,t}$ and $\widehat{\bm{C}}$, and define $\widehat{\bm{\Sigma}}_{t_1,\ldots,t_J}=\big[\widehat{\bm{C}}(t_a,t_b)\big]_{a,b=1}^J$.

\Statex \textbf{Synthetic data generation:}
\For{$f=1,\ldots,F$}
    \State Sample $\bm{z}^{(f)}_{1:J}:=\big(\bm{z}^{(f)}(t_1)^\top,\ldots,\bm{z}^{(f)}(t_J)^\top\big)^\top \sim N(\bm{0},\widehat{\bm{\Sigma}}_{t_1,\ldots,t_J})$.
    \For{$j=1,\ldots,J$}
        \State Solve $\frac{\mathrm{d}}{\mathrm{d}u}\bm{x}^{(f)}_{u}(t_j)=\widehat{\bm{v}}_{u,t_j}(\bm{x}^{(f)}_{u}(t_j))$ with $\bm{x}^{(f)}_{0}(t_j)=\bm{z}^{(f)}(t_j)$, and set $\tilde{\bm{X}}^{(f)}(t_j)=\bm{x}^{(f)}_{1}(t_j)$.
    \EndFor
\EndFor

\Statex \textbf{Data completion:}
\If{$i^\star$ is specified}
    \State Compute pullback samples $\widehat{\bm{Z}}_{i^\star j}$ from $\bm{Y}_{i^\star j}$ via \eqref{pull_back_sample}, and stack them as
    $\bm{z}^{\mathrm{obs}}_{i^\star}:=(\widehat{\bm{Z}}_{i^\star 1}^\top,\ldots,\widehat{\bm{Z}}_{i^\star J_{i^\star}}^\top)^\top$. Let $\bm{Z}_{\mathrm{obs}}:=\big(\bm{Z}(T_{i^\star 1})^\top,\ldots,\bm{Z}(T_{i^\star J_{i^\star}})^\top\big)^\top$, and let $\bm{Z}_{\mathrm{mis}}$ be the latent variables at $\{t_1,\ldots,t_{J'}\}:=\{t_1,\ldots,t_J\}\setminus\{T_{i^\star 1},\ldots,T_{i^\star J_{i^\star}}\}$.
    \State Partition the Gaussian law induced by $\widehat{\bm{C}}$ as
    \(
    \begin{pmatrix}\bm Z_{\mathrm{obs}}\\ \bm Z_{\mathrm{mis}}\end{pmatrix}
    \sim
    N\!\left(
    \bm 0,\
    \begin{pmatrix}
    \widehat{\bm{\Sigma}}_{\mathrm{obs},\mathrm{obs}} & \widehat{\bm{\Sigma}}_{\mathrm{obs},\mathrm{mis}}\\
    \widehat{\bm{\Sigma}}_{\mathrm{mis},\mathrm{obs}} & \widehat{\bm{\Sigma}}_{\mathrm{mis},\mathrm{mis}}
    \end{pmatrix}
    \right),
    \)
    and compute
    \(
    \widehat{\bm{\mu}}_{\mathrm{mis}\mid\mathrm{obs}}
    =
    \widehat{\bm{\Sigma}}_{\mathrm{mis},\mathrm{obs}}
    \widehat{\bm{\Sigma}}_{\mathrm{obs},\mathrm{obs}}^{-1}
    \bm z^{\mathrm{obs}}_{i^\star},\
    \widehat{\bm{\Sigma}}_{\mathrm{mis}\mid\mathrm{obs}}
    =
    \widehat{\bm{\Sigma}}_{\mathrm{mis},\mathrm{mis}}
    -
    \widehat{\bm{\Sigma}}_{\mathrm{mis},\mathrm{obs}}
    \widehat{\bm{\Sigma}}_{\mathrm{obs},\mathrm{obs}}^{-1}
    \widehat{\bm{\Sigma}}_{\mathrm{obs},\mathrm{mis}}.
    \)
    \For{$f=1,\ldots,F$}
     \State Sample $\bm z^{(f)}_{\mathrm{mis}}\sim N(\widehat{\bm{\mu}}_{\mathrm{mis}\mid\mathrm{obs}},\widehat{\bm{\Sigma}}_{\mathrm{mis}\mid\mathrm{obs}})$; let
\(
\big(\bm z^{(f)}_{\mathrm{mis}}(t_1)^\top,\ldots,\bm z^{(f)}_{\mathrm{mis}}(t_{J'})^\top\big)^\top=\bm z^{(f)}_{\mathrm{mis}}
\);
and for each $j=1,\ldots,J'$, solve
$\frac{\mathrm{d}}{\mathrm{d}u}\bm{x}^{(f)}_{u}(t_j)=\widehat{\bm{v}}_{u,t_j}(\bm{x}^{(f)}_{u}(t_j))$
with $\bm{x}^{(f)}_{0}(t_j)=\bm{z}^{(f)}_{\mathrm{mis}}(t_j)$; set $\tilde{\bm X}^{(f)}_{i^\star}(t_j)=\bm{x}^{(f)}_{1}(t_j)$.
    \EndFor
\EndIf

\State Output the synthetic trajectories and, if applicable, the completion samples.
\end{algorithmic}
\end{algorithm}

For data synthesis, we sample $\bm{z}_{1:J}^{(f)}$ from $p_{\bm{Z},t_1,\ldots,t_J}$, where $p_{\bm{Z},t_1,\ldots,t_J}$ is the mean-zero Gaussian distribution induced by the estimated kernel $\widehat{\bm{C}}$ from Algorithm~\ref{alg:pfm_full}. For data completion for subject $i$ on a time grid, we sample $\bm{z}_{1:J}^{(f)}$ conditional on the observed data $\bm{Y}_{ij}$, $j=1,\ldots,J_i$. To achieve this, we use the pullback samples $\hat{\bm{Z}}_{ij}$, $j=1,\ldots,J_i$, from \eqref{pull_back_sample}, and then sample $\bm{z}_{1:J}^{(f)}$ from the Gaussian distribution conditional on these pullback samples, where the Gaussian distribution is determined by $\widehat{\bm{C}}$.
The resulting $\{\tilde{\bm{X}}^{(f)}(t_j): j=1,\ldots,J\}$, $f=1,\ldots,F$, are longitudinal samples over the time grid conditional on $\bm{Y}_{ij}$, $j=1,\ldots,J_i$. We summarize data synthesis and completion in Algorithm~\ref{alg:pfm_synthetic}.

A valuable feature of Algorithm~\ref{alg:pfm_synthetic} is that it is trained directly on irregular longitudinal data, while generation/completion are carried out through an expressive continuous transport framework on any user-specified time grid.
In contrast, existing continuous generative methods, such as diffusion models \citep{ho2020denoising,yang2023diffusion} and
standard flow matching \citep{liu2022flow,lipman2022flow}, are typically developed for fixed grids and may require additional preprocessing to accommodate irregular panels.

\begin{algorithm}[ht!]
\caption{Longitudinal Classification via Panel Flow Matching}
\label{alg:pfm_classification}
\begin{algorithmic}[1]
\Require Labeled panel $\{(G_i,\{(T_{ij},\bm{Y}_{ij})\}_{j=1}^{J_i}): i=1,\ldots,n\}$; new panel $\{(T^{\mathrm{new}}_{ij},\bm{Y}^{\mathrm{new}}_{ij})\}_{j=1}^{J_i}$; prior probabilities $\pi_0$ and $\pi_1$, or depth $L$, width $m$, and synthetic sample size $F$;
\Ensure Predicted label $\widehat{G}_i$.

\Statex \textbf{Stage I: Train class-conditional panel flow models}
\For{$g\in\{0,1\}$}
    \State Form class-$g$ training set $\mathcal{D}_g=\{(T_{ij},\bm{Y}_{ij}): G_i=g,\ j=1,\ldots,J_i\}$.
    \State Apply Algorithm~\ref{alg:pfm_full} to $\mathcal{D}_g$ to obtain $(\widehat{\bm{v}}^{(g)}_{u,t},\widehat{\bm{C}}^{(g)})$.
\EndFor

\Statex \textbf{Stage II (Optional): Bayesian classification}
\State Compute, for each $g\in\{0,1\}$,
\[
\widehat{\ell}_g(i)
:=
\log \widehat{p}^{(g)}_{\bm{X},T^{\mathrm{new}}_{i1},\ldots,T^{\mathrm{new}}_{iJ_i}}
\!\big(\bm{Y}^{\mathrm{new}}_{i1},\ldots,\bm{Y}^{\mathrm{new}}_{iJ_i}\big)
+\log \pi_g,
\]
where $\widehat{p}^{(g)}_{\bm{X},T^{\mathrm{new}}_{i1},\ldots,T^{\mathrm{new}}_{iJ_i}}$ is evaluated via
\eqref{joint_dis} with $\bm{v}_{u,t}$ replaced by $\widehat{\bm{v}}^{(g)}_{u,t}$ and the latent joint
density taken as the mean-zero Gaussian distribution induced by $\widehat{\bm{C}}^{(g)}$.
\State Output $\widehat{G}^{\mathrm{Bayes}}_i:=\arg\max_{g\in\{0,1\}}\widehat{\ell}_g(i)$.

\Statex \textbf{Stage II (Optional): Synthetic classification}
\For{$g\in\{0,1\}$}
    \For{$f=1,\ldots,F$}
        \State Sample latent panel $\bm{z}^{(f)}_{g,1:J_i}\sim p^{(g)}_{\bm{Z},T_{i1}^{\mathrm{new}},\ldots,T_{iJ_i}^{\mathrm{new}}}$ induced by $\widehat{\bm{C}}^{(g)}$.
        \State For $j=1,\ldots,J_i$, solve
        $\frac{\mathrm{d}}{\mathrm{d}u}\bm{x}_{u}(T_{ij}^{\mathrm{new}})=\widehat{\bm{v}}^{(g)}_{u,T_{ij}^{\mathrm{new}}}\big(\bm{x}_{u}(T_{ij}^{\mathrm{new}})\big)$
        with $\bm{x}_0(T_{ij}^{\mathrm{new}})=\bm{z}^{(f)}_{g,j}$ and set $\tilde{\bm{X}}^{(f)}_{g}(T_{ij}^{\mathrm{new}})=\bm{x}_1(T_{ij}^{\mathrm{new}})$.
        \State Set $\bm{s}^{(f)}_{g}:=\mathrm{vec}\{\tilde{\bm{X}}^{(f)}_{g}(T_{i1}^{\mathrm{new}}),\ldots,\tilde{\bm{X}}^{(f)}_{g}(T_{iJ_i}^{\mathrm{new}})\}\in\mathbb{R}^{pJ_i}$ and label $\tilde{G}^{(f)}_{g}:=g$.
    \EndFor
\EndFor
\State Construct synthetic training set $\widetilde{\mathcal{D}}=\{(\bm{s}^{(f)}_{g},\tilde{G}^{(f)}_{g}): g\in\{0,1\},\, f=1,\ldots,F\}$.
\State Train a classifier $\widehat{h}:\mathbb{R}^{pJ_i}\to\mathbb{R}$ in $\mathcal{F}_{pJ_i,1}(L,m)$ on $\widetilde{\mathcal{D}}$ by minimizing the logistic loss
\[
\widehat{\mathcal{L}}_{\mathrm{cla}}(h)
=
\frac{1}{2F}\sum_{g\in\{0,1\}}\sum_{f=1}^{F}
\Big[
-\tilde{G}^{(f)}_{g}\,h\!\big(\bm{s}^{(f)}_{g}\big)
+\log\!\big\{1+\exp\big(h(\bm{s}^{(f)}_{g})\big)\big\}
\Big].
\]
\State Output
$\widehat{G}^{\mathrm{Synthe}}_i:=\mathbb{I}\{\widehat{h}(\bm{s}^{\mathrm{new}})\ge 0\}$ with
$\bm{s}^{\mathrm{new}}:=\mathrm{vec}\{\bm{Y}^{\mathrm{new}}_{i1},\ldots,\bm{Y}^{\mathrm{new}}_{iJ_i}\}$.
\end{algorithmic}
\end{algorithm}

\subsection{Longitudinal Classification}
Algorithm~\ref{alg:pfm_full} provides a flexible tool for longitudinal classification. Suppose
each subject $i$ is associated with a class label $G_i\in\{0,1\}$, and the sample size for class $g$ is denoted by $n_g$, $g\in \{0,1\}$, with $n_0+n_1=n$. A
direct strategy is to learn class-conditional panel flows by applying Algorithm~\ref{alg:pfm_full}
separately to the subsets of subjects in each class, yielding estimated vector fields
$\widehat{\bm{v}}^{(g)}_{u,t}$ and latent kernels $\widehat{\bm{C}}^{(g)}$ for
$g\in\{0,1\}$. In the following, we use the estimated vector fields and kernels for classification.

\paragraph*{Bayesian Viewpoint.}
For a new subject with observed panel
$\{(T^{\mathrm{new}}_{ij},\bm{Y}^{\mathrm{new}}_{ij})\}_{j=1}^{J_i}$, we predict their class label by computing a class score based on the panel log-posterior
\[
\widehat{\ell}_g(i)
:=
\log \widehat{p}^{(g)}_{\bm{X},T^{\mathrm{new}}_{i1},\ldots,T^{\mathrm{new}}_{iJ_i}}
\!\big(\bm{Y}^{\mathrm{new}}_{i1},\ldots,\bm{Y}^{\mathrm{new}}_{iJ_i}\big)
+\log \pi_g,
\]
where $\pi_g$ is the prior probability of class $g$, and
$\widehat{p}^{(g)}_{\bm{X},T^{\mathrm{new}}_{i1},\ldots,T^{\mathrm{new}}_{iJ_i}}$ is evaluated via
\eqref{joint_dis}, with $\bm{v}_{u,t}$ replaced by $\widehat{\bm{v}}^{(g)}_{u,t}$ for computing the inverse maps and their Jacobians, and with the latent joint density taken to be the mean-zero Gaussian distribution induced by $\widehat{\bm{C}}^{(g)}$.
The predicted label is then determined by Bayes' rule:
\(
\widehat{G}_i=\arg\max_{g\in\{0,1\}}\widehat{\ell}_g(i).
\)

An advantage of the above procedure is that it avoids a dimension-reduction step such as \eqref{KL_expansion} for longitudinal classification. Such steps are usually required by existing methods \citep{muller2005functional,wang2024review,xue2024optimal}, which project both training and testing trajectories onto some unified basis functions while ignoring the class label. This procedure may discard discriminative information and introduce dimension-reduction error in both training and prediction.

\paragraph*{Synthetic Viewpoint.}
Our method also provides a synthetic procedure for classification.
Specifically, for a pre-specified grid
$T^{\mathrm{new}}_{i1},\ldots,T^{\mathrm{new}}_{iJ_i}$ for a subject, we generate synthetic
trajectories from each class-conditional model and use the resulting vectors
\begin{eqnarray}\label{syn_data}
    \mathrm{vec}\big\{\tilde{\bm{X}}^{(f)}_g(T^{\mathrm{new}}_{i1}),\ldots,\tilde{\bm{X}}^{(f)}_g(T^{\mathrm{new}}_{iJ_i})\big\}
\in\mathbb{R}^{pJ_i},\qquad f=1,\ldots,F,\ \ g\in\{0,1\},
\end{eqnarray}
as training samples for a classifier. Here,
$\{\tilde{\bm{X}}^{(f)}_g(T^{\mathrm{new}}_{ij})\}_{j=1}^{J_i}$ is generated by first sampling a
latent panel from the Gaussian distribution induced by $\widehat{\bm{C}}^{(g)}$, and then pushing it
forward via the class-specific vector field $\widehat{\bm{v}}^{(g)}_{u,t}$ by solving the forward
ODE, as in Algorithm~\ref{alg:pfm_synthetic}. One benefit of the synthetic approach is that it can mitigate class imbalance in the raw data (e.g.,
$n_0\ll n_1$) by generating an equal number of synthetic samples from each class for classifier
training. This
procedure also avoids dimension reduction of longitudinal data at prediction time, as the
synthetic data are time-grid adaptive for a new subject.

As we can generate an arbitrarily large amount of synthetic data in \eqref{syn_data} (i.e., large $F$), we can employ flexible vector-based methods for classification training. Here, we focus on neural-network classification by training a multilayer perceptron on the synthetic data.

We summarize both the Bayesian-rule-based and synthetic approaches for longitudinal classification
in Algorithm~\ref{alg:pfm_classification}. Similar to Algorithm~\ref{alg:pfm_full}, we
employ denoising on the training and testing data before model fitting, and set $L=3$ and $m=30$
for the classification network. In our implementation, we take the synthetic sample size to
$F=2000$.

\section{Simulation}\label{sec:sim}

We consider multivariate longitudinal data on a fixed time domain
$\mathcal{T}=[0,1]$. For each subject
$i=1,\ldots,n$, we draw the number of visits $J_i$ and then sample observation times $T_{ij} \stackrel{\mathrm{iid}}{\sim} \mathrm{Unif}(\mathcal{T})$, $j=1,\ldots,J_i$.
Let $\bm{X}_i(t):=(X_{i1}(t),\ldots,X_{ip}(t))^\top\in\mathbb{R}^p$ be an underlying $p$-feature stochastic process.
To generate $\bm{X}_i(\cdot)$, we use the model
\begin{equation}\label{model_dat}
\bm{Y}_{ij} =\bm{X}_i(T_{ij})+\bm{\varepsilon}_{ij}= \bm{\mu}(T_{ij}) + \sum_{k=1}^{20} \xi_{ik}\,\bm{\psi}_k(T_{ij})+\bm{\varepsilon}_{ij},\qquad i = 1,\ldots,n,\ j=1,\ldots,J_i,
\end{equation}
where $T_{ij}$ are uniformly distributed time points on $\mathcal{T}$, $\bm{\mu}(t)\in\mathbb{R}^p$ is the mean function, $\bm{\psi}_k(t)\in\mathbb{R}^p$ are
vector-valued functions, the scores are independent across $k$ with
$\xi_{ik} \sim \mathcal{N}(0,\exp(-(k-1)))$, $k=1,\ldots,20$, and $\bm{\varepsilon}_{ij}$ is mean-zero Gaussian noise independent across $i$, $j$, and different features, with standard deviation set as $0.05\cdot \bm{X}_i(T_{ij})$. The generation of $\bm{\mu}$ and $\bm{\psi}_k$ is detailed in Part~\ref{sec:sim_SM} of the Supplementary Materials. 
In addition, we consider a non-Gaussian case by modifying the distribution of $\xi_{ik}$ to follow a centered Gamma distribution with shape parameter $2$ and scale
\(
\sqrt{\frac{\exp\big(-(k-1)\big)}{2}},
\)
which yields $\mathrm{Var}(\xi_{ik})=\exp\big(-(k-1)\big)$.

Let $p_t$ denote the true panel density of $\bm{X}_i(t)$, and let $\widehat p_t$ be an estimator constructed from the observed data. We use the integrated squared error
\begin{equation*}
\mathrm{ISE}
=
\int_{0.05}^{0.95}\int_{[-3,3]^p}\big| \log\widehat p_t(\bm{x})-\log p_t(\bm{x})\big|p_t(\bm{x})\,\mathrm{d}\bm{x}\,\mathrm{d}t,
\end{equation*}
where $[-3,3]^p$ covers the high-density region of $p_t$ under our setting.
In implementation, we approximate $\mathrm{ISE}$ on a $10^{p+1}$ rectangular grid over $[0.05,0.95]\times[-3,3]^p$.

Furthermore, we consider two additional tasks to evaluate full density estimation for longitudinal data: (i) trajectory generation and (ii) trajectory classification.
For trajectory generation, we generate data from the same Gaussian and non-Gaussian settings as above and fit data generators to the resulting longitudinal dataset. We then quantify generation accuracy by the population $2$-Wasserstein distance between the distribution of a true trajectory $\bm X(\cdot)$ and that of a generated trajectory $\widetilde{\bm X}(\cdot)$ produced by the fitted data generators:
\begin{equation*}
W_2\!\left(\bm X,\widetilde{\bm X}\right)
=
\left\{
\inf_{\gamma\in\Gamma(\bm X,\,\widetilde{\bm X})}
\mathbb{E}_{(\bm X,\widetilde{\bm X})\sim\gamma}\big\|\bm X-\widetilde{\bm X}\big\|_2^2
\right\}^{1/2},
\end{equation*}
where $\Gamma(\bm X,\widetilde{\bm X})$ is the set of all couplings of the two distributions. In implementation, we draw $1000$ samples from each distribution and compute the distance using the \texttt{transport} package in \texttt{R} \citep{gottschlich2014shortlist}.

For trajectory classification, we generate a subject-level label $Y_i \sim \mathrm{Bernoulli}(1/2)$ and then generate a $p$-dimensional trajectory $\bm X_i(\cdot)$ at irregular times $\{T_{ij}\}_{j=1}^{J_i}$. 
When $Y_i=0$, we generate the longitudinal data from the baseline Gaussian or non-Gaussian setting used above. 
When $Y_i=1$, we keep the same sampling design and the same score family, but perturb the mean and covariance structures of the data; see Part~\ref{sec:sim_SM} of the Supplementary Materials for the data-generation details.
Classification accuracy is evaluated by the prediction accuracy
\[
\frac{1}{100}\sum_{i=1}^{100}\mathbb{I}(\widehat Y_i^{\mathrm{test}}=Y_i^{\mathrm{test}}),
\]
where $Y_i^{\mathrm{test}}$ is the test label and $\widehat Y_i^{\mathrm{test}}$ is the predicted label based on the longitudinal data in the test set.

We vary the sample size $n$ and the visit counts $J_i\sim \mathrm{Unif}\{4,\ldots,8\}$. For panel density evaluation, we set $p=2$ and $4$, while for classification and generation, we set $p=10$. All simulations are replicated 100 times.

\noindent\textbf{Comparison Methods.}
We compare our method with several baseline approaches across the three evaluation tasks.
For panel density estimation, we consider kernel density estimation \citep[KDE;][]{chen2017tutorial}. Since KDE is applicable only to regular-grid representations, we first reconstruct each subject's trajectories using FPCA \citep{yao2005functional}, i.e., we estimate the mean function, eigenfunctions, and scores, and then reconstruct the corresponding curves. We then apply KDE to the reconstructed samples.

For trajectory generation, our method is directly applicable to irregular longitudinal data and generates longitudinal data for any provided time grid (see Algorithm~\ref{alg:pfm_synthetic}).
For comparison, we apply a variational autoencoder \citep[VAE;][]{ramchandran2021longitudinal} with latent dimension set as $p$; this approach accommodates irregular longitudinal data similarly to our method.
We further include a diffusion model \citep[DM;][]{ho2020denoising} and a standard flow-matching method \citep[FM;][]{lipman2022flow} as baselines, which we implement in the FPCA score space since these methods require vector inputs. Specifically, we train DM/FM to generate new score vectors, and then map the generated scores back to trajectories by combining them with the estimated FPCA mean and eigenfunctions. These longitudinal generation methods are evaluated on an equally spaced 20-point grid in $[0,1]$.

For trajectory classification, we consider PFM under the Bayesian and synthetic frameworks, as in Algorithm~\ref{alg:pfm_classification}, denoted by PFM-Bayes and PFM-Synthe, respectively. Meanwhile, VAE can be used for trajectory classification by forming a class probability based on the variational approximation \citep{ramchandran2021longitudinal}.
In addition, we include standard tabular classifiers---logistic regression (LR), linear discriminant analysis (LDA), and a neural network (NN)---which require vector inputs. We again apply FPCA and perform classification on the extracted training score vectors, and predict test-set labels using the corresponding test scores. 
For all the above methods, we assign a class label to a new longitudinal sample by selecting the class with the larger estimated posterior probability. For PFM-Bayes, VAE, and LDA, the prior class probabilities used in the posterior probability calculation are set to the corresponding class proportions in the training sample.

For the neural-network-based methods (PFM, VAE, NN, DM, and FM), we use the same basic architecture: a multilayer perceptron with depth $3$ and width $30$. The FPCA step used for generation and classification is detailed in Part~\ref{sec:FPCA} of the Supplementary Materials.

\begin{table}[!h]
\renewcommand{\arraystretch}{1.1}
\setlength\tabcolsep{8pt}
\centering
\caption{Mean ISEs over 100 simulation replications.}
\label{tab:mse_grouped_gaussian_nongaussian}
\begin{tabular}{c|cc|cc|cc|cc}
\hline
& \multicolumn{4}{c|}{Gaussian} & \multicolumn{4}{c}{Non-Gaussian} \\
\cline{2-9}
& \multicolumn{2}{c|}{$p=2$} & \multicolumn{2}{c|}{$p=4$}
& \multicolumn{2}{c|}{$p=2$} & \multicolumn{2}{c}{$p=4$} \\
$n$ & KDE & PFM & KDE & PFM & KDE & PFM & KDE & PFM \\
\hline
100 & 0.2800 & 0.0306 & 0.0826 & 0.0012 & 0.5195 & 0.0689 & 7.9130 & 0.3569 \\
200 & 0.2401 & 0.0143 & 0.0470 & 0.0009 & 0.3819 & 0.0374 & 4.8342 & 0.2600 \\
300 & 0.1620 & 0.0113 & 0.0377 & 0.0008 & 0.3308 & 0.0296 & 4.1025 & 0.2362 \\
\hline
\end{tabular}
\end{table}

\begin{figure}[ht!]
    \centering
    \includegraphics[width=0.9\linewidth]{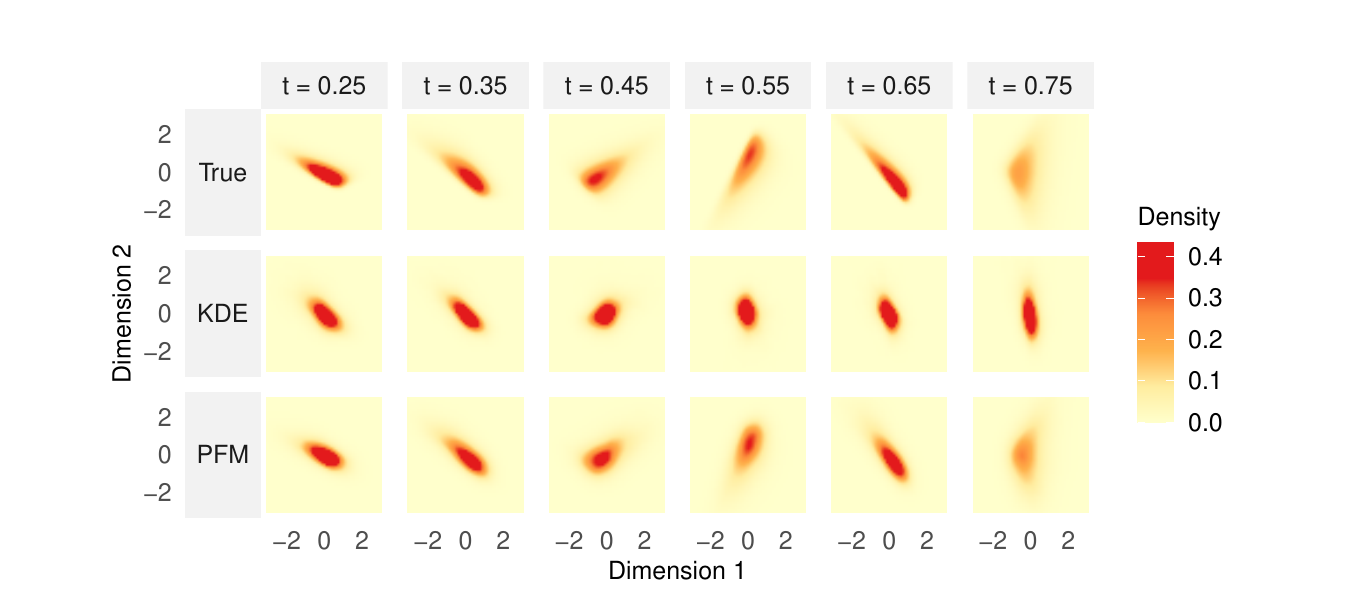}
    \caption{One true panel density and its estimated panel densities averaged over 100 simulations for $p=2$ and $n=200$. The simulations are conducted under the non-Gaussian setting.}
    \label{fig:PDE_sim}
\end{figure}

\noindent\textbf{Results.} The ISEs from 100 simulation replications are reported in Table~\ref{tab:mse_grouped_gaussian_nongaussian}. We observe that our method becomes more accurate as $n$ increases in each setting, consistent with the statistical convergence in Theorem~\ref{thm:pfm_rate_like32}. 
In contrast, KDE yields substantially larger ISEs, and the ratio \(\text{KDE}/\text{PFM}\) for the mean ISE becomes significantly larger for \(p=4\) (see Table~\ref{SM_tab_ratio} in the Supplementary Materials for the ratio values). This is likely because the initial FPCA reconstruction introduces dimension-reduction error, and KDE is generally less flexible for higher-dimensional density estimation.
We also illustrate one true non-Gaussian panel density and its estimates in Figure~\ref{fig:PDE_sim} for $p=2$.  
We observe that KDE only partially captures the high-density region, because the FPCA preprocessing introduces model bias. Meanwhile, PFM better captures the time-varying density shape than KDE.

\begin{figure}[ht!]
    \centering
    \includegraphics[width=0.7\linewidth]{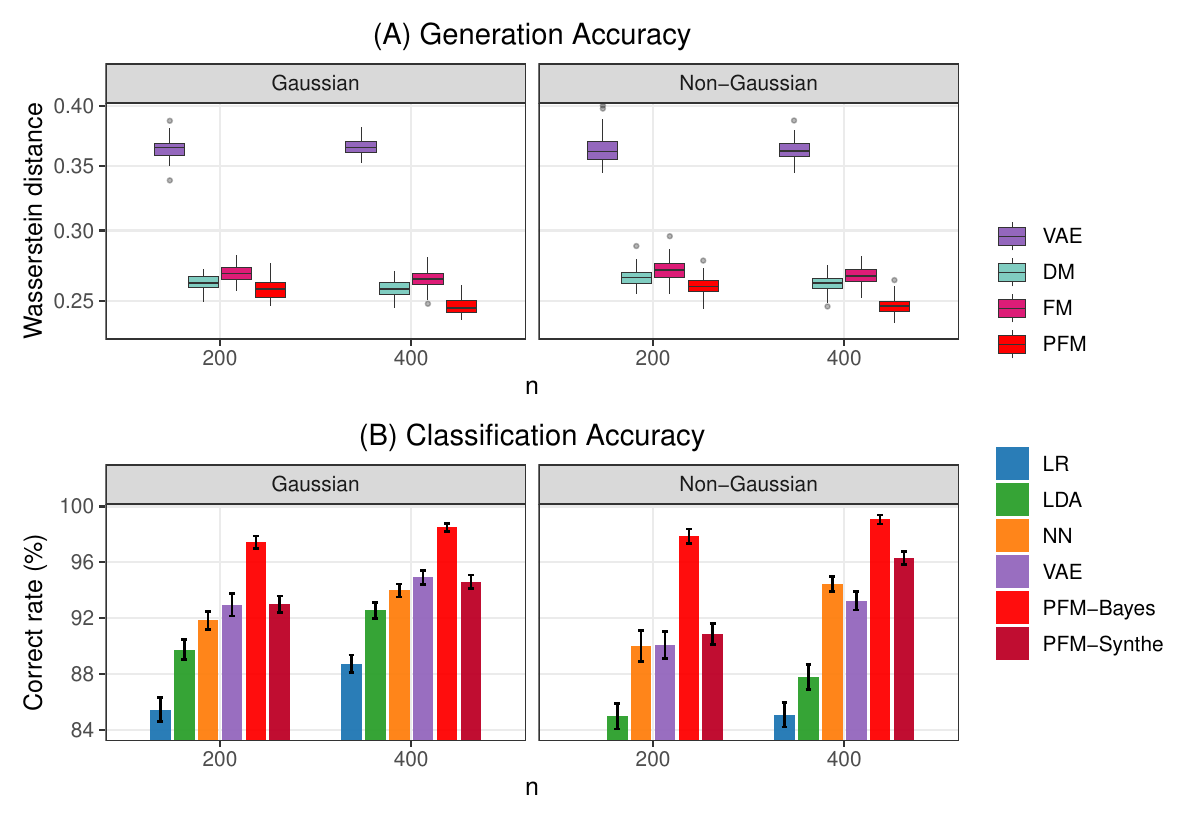}
    \caption{\textbf{(A)} The Wasserstein distance for trajectory generation and \textbf{(B)} the prediction accuracy for trajectory classification.}
    \label{fig:cla_gen_sim}
\end{figure}

The generation accuracy is shown in Figure~\ref{fig:cla_gen_sim}(A). We observe that VAE performs the worst, likely due to its limited expressive power. DM and FM are more flexible generative methods, but they can only be implemented on the FPCA scores and thus suffer from dimension-reduction error. In contrast, our method directly handles irregular longitudinal data, leading to better overall performance. A similar pattern is also observed for classification; see Figure~\ref{fig:cla_gen_sim}(B). In both the Gaussian and non-Gaussian settings, VAE is consistently inferior to PFM. The two-step baselines (LR, LDA, and NN) mostly underperform PFM-Bayes and PFM-Synthe, especially in the non-Gaussian setting, likely because FPCA introduces error through dimension reduction and Gaussian assumptions. We also observe that PFM-Synthe is inferior to PFM-Bayes, as PFM-Synthe may introduce additional error due to the finite number of generated samples. 

\section{Real Data Analysis}\label{sec:real}

We analyze the vaginal microbiome longitudinal dataset of \citet{costello2022vaginal} archived in the Stanford Digital Repository (\url{https://purl.stanford.edu/pz745bc9128}). The dataset consists of repeated vaginal swab samples collected longitudinally for each pregnancy at irregular, pregnancy-specific time points during gestation. The dataset comprises 196 pregnancies and 3848 vaginal swab samples, each represented by a vector of counts measured over microbial features.

We preprocess the data as follows. We first retain pregnancies with longitudinal follow-up by requiring at least four swabs per pregnancy, and we focus on mid-to-late gestation by keeping samples collected between 12 and 37 gestational weeks. Starting from the raw count table, we aggregate microbial features to the genus level and transform counts into within-sample relative abundances by adding a small pseudocount (0.5) to each genus and normalizing by the sample total. We then retain genera whose relative abundance exceeds \(5\times 10^{-4}\) and that are observed in at least one-third of all pregnancies, and take logarithms of the resulting relative abundances. After preprocessing, we obtain an irregularly sampled longitudinal dataset with sample size \(n=188\) pregnancies and a total of \(\sum_{i=1}^n J_i = 2820\) observed time points, each represented by a vector of \(21\) genus-level features.

\paragraph*{Longitudinal Classification.}
In the resulting cohort, each pregnancy is assigned a binary outcome label (preterm vs.\ term), with \(47\) pregnancies labeled as preterm and the remaining \(141\) pregnancies labeled as term.
Understanding pregnancy status from vaginal microbiome measurements is important for characterizing pregnancy-related risks and for early identification of preterm risk.
We therefore focus on estimating a classifier for pregnancy status using the vaginal swab longitudinal trajectories during gestation.

To mitigate correlation among microbiome features, we first cluster the 21 genera based on
Spearman correlation. Specifically, we compute pairwise Spearman correlations across the 21 genera
using the log-transformed relative abundances from the 2,820 time points, and then
perform hierarchical clustering based on the $(1-|\text{correlation}|)$ matrix to group the genera.
The resulting dendrogram is given in Figure~\ref{fig:dendrogram} in the Supplementary Materials. We
observe that one genus, \textit{Ureaplasma}, forms a separate cluster, while the remaining genera
are grouped into two clusters.

\begin{figure}[h]
    \centering
    \includegraphics[width=0.9\linewidth]{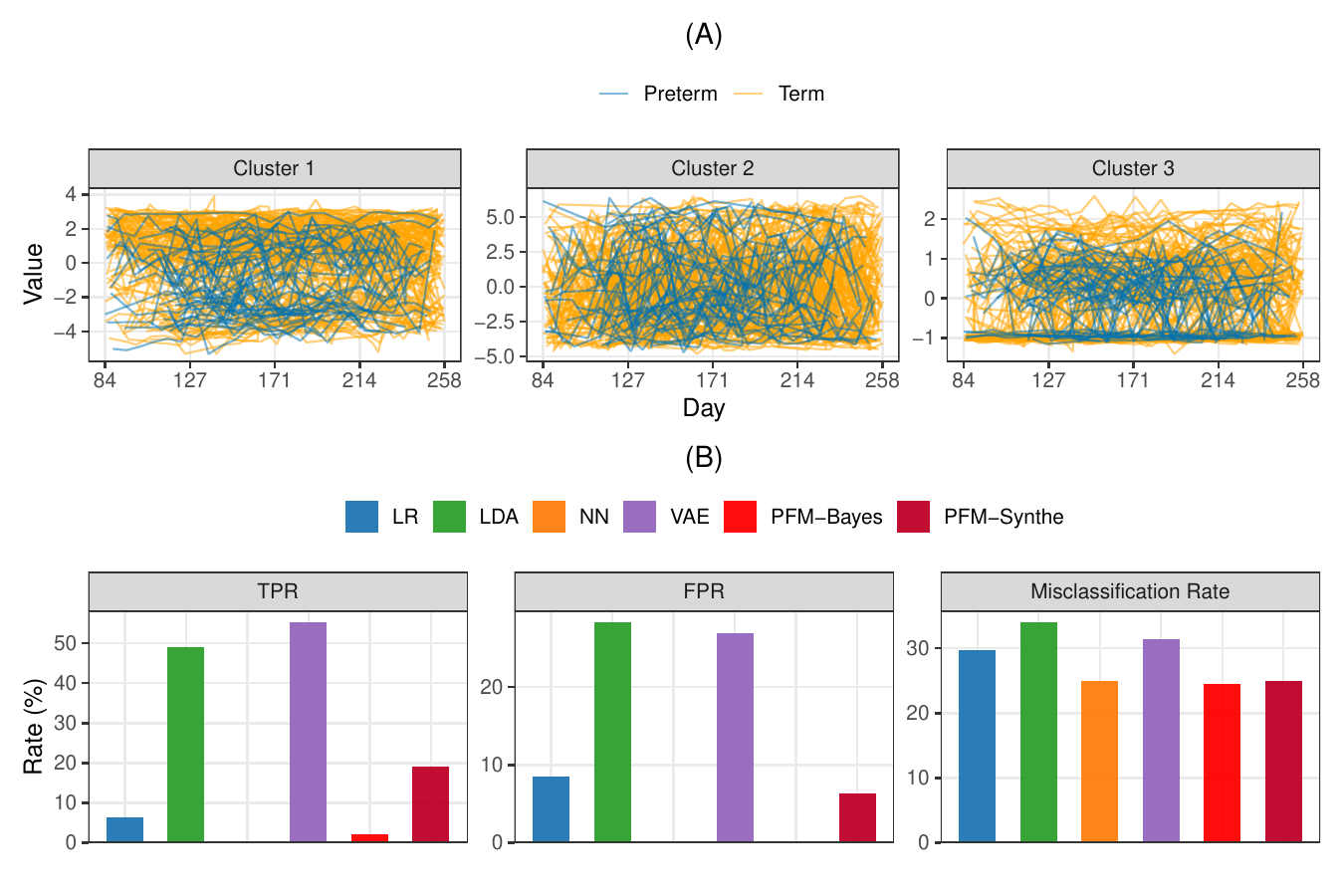}
    \caption{(A) Observed longitudinal principal component (PC) score trajectories. (B) Leave-one-out cross-validation performance for longitudinal classification (term vs.\ preterm), summarized by the true positive rate (TPR), false positive rate (FPR), and misclassification rate.}
    \label{fig:dat_illu_ge}
\end{figure}

We set the number of clusters to 3 to prune the dendrogram. 
Accordingly, we conduct principal component analysis (PCA) on the data across the \(2820\) time points within each cluster. We take the first principal component from each cluster, which explains \(62\%\), \(65\%\), and \(100\%\) of the variance, respectively. Let \(Y_i\in\{0,1\}\) be the pregnancy outcome label for pregnancy \(i\), with \(Y_i=1\) indicating preterm and \(Y_i=0\) indicating term. The resulting longitudinal cluster-PC score trajectories for different pregnancy outcomes are shown in Figure~\ref{fig:dat_illu_ge}(A). We see that the term and preterm groups exhibit substantial overlap in several components, and that the sampling schedules are highly irregular across pregnancies.

We apply the classification methods considered in our simulation study to this dataset, including LR, LDA, NN, VAE, PFM-Bayes, and PFM-Synthe, where the prior probabilities for LDA, VAE, and PFM-Bayes are set to \(0.5\) for the two classes.
To compare different methods, we use leave-one-out cross-validation: for each held-out pregnancy, we train the different methods on the remaining \((n-1)\) pregnancies and predict the held-out label as in the simulation study. We summarize classification performance using (i) the true positive rate (TPR, the proportion of preterm pregnancies correctly classified as preterm), (ii) the false positive rate (FPR, the proportion of term pregnancies incorrectly classified as preterm), and (iii) the misclassification rate (the proportion of incorrectly classified pregnancies among all pregnancies). The results are reported in Figure~\ref{fig:dat_illu_ge}(B).

We find that LDA and VAE achieve higher TPR for identifying preterm pregnancies, but this improvement comes at the cost of substantially higher FPR. Conversely, LR, NN, and PFM-Bayes exhibit lower TPR but markedly smaller FPR, indicating a tendency to classify pregnancies as term. The results for LR, NN, and PFM-Bayes may be influenced by the class imbalance in the analysis cohort (\(47\) preterm vs.\ \(141\) term pregnancies), under which these methods tend to favor the dominant class and thus yield low TPR. In contrast, LDA and VAE appear to underfit the dominant class, as they use relatively restrictive modeling for longitudinal data. This leads to higher TPR but also inflated FPR.

Overall, PFM-Synthe attains an overall accuracy similar to that of NN and PFM-Bayes and outperforms the other methods.
In particular, PFM-Synthe provides a favorable trade-off:
it achieves a higher TPR than LR, NN, and PFM-Bayes, while maintaining a smaller FPR than LDA and
VAE. 
This improvement is consistent
with the use of synthetic trajectory augmentation, which helps mitigate class imbalance by generating
balanced, time-grid-adaptive training data for both the minority (preterm) and majority (term) classes.

\begin{figure}[ht!]
    \centering
    \includegraphics[width=0.9\linewidth]{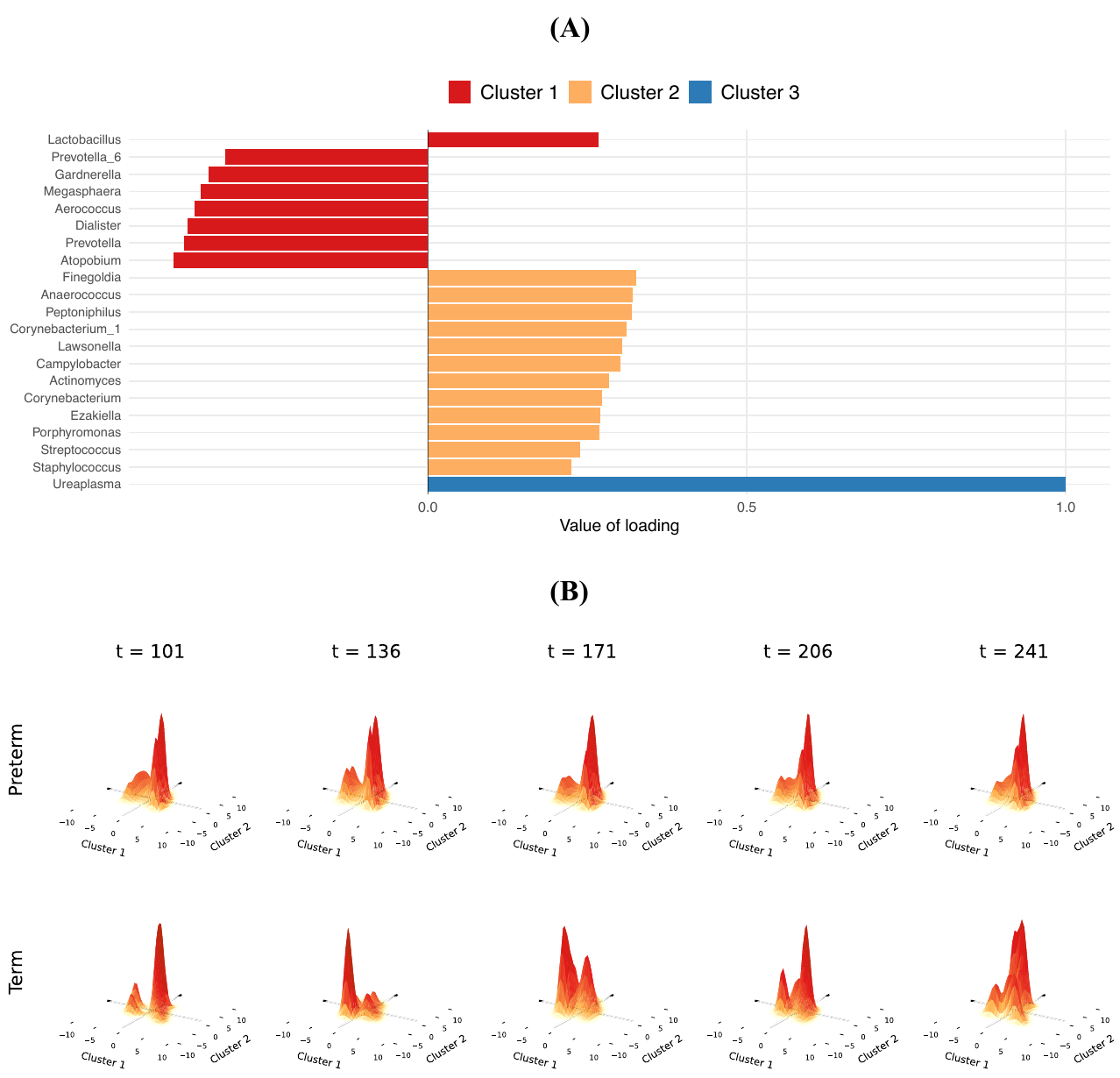}
    \caption{\textbf{(A)} The first PC loadings for each cluster and \textbf{(B)} PFM-based panel density estimates for the cluster-PC trajectories for preterm vs.\ term pregnancy.}
    \label{fig:panel_g}
\end{figure}

\paragraph*{Panel Density Estimation.}
To illustrate differences between the preterm and term groups, we display the PFM-based estimates of the panel density for the vaginal swab longitudinal trajectories.
For illustration, we focus on the longitudinal PCs from the first two clusters, which contain information on the 20 genera other than \textit{Ureaplasma}. The corresponding PC loadings and panel density estimates are shown in Figure~\ref{fig:panel_g}.

By the loading values of the clusters, we observe that Cluster~1 primarily contrasts a \textit{Lactobacillus}-enriched profile against a collection of genera with negative loadings. 
In contrast, Cluster~2 has positive loadings for other taxa. 
The estimated densities of these two clusters reveal time-varying patterns with clearly different concentration regions between the preterm and term cohorts, as well as pronounced non-Gaussian features. Notably, the preterm panel density appears comparatively stable over time, whereas the term density evolves more substantially as gestation progresses.

For the preterm group, the panel density places most of its probability mass in a central region across all times, with a persistent mode at mildly positive Cluster~1/Cluster~2 values. In contrast, the term group shows much more pronounced temporal variation: around $t=101$ and $t=171$, the density shifts clearly toward negative Cluster~1 values, while at later times it moves back toward a more central configuration, with high mass placed on positive Cluster~2 values. These results suggest that temporal shifts in \textit{Lactobacillus} and the other genera in Cluster~1 may provide a useful signal for distinguishing term from preterm pregnancy.

\section{Discussion}\label{sec:dis}

In this article, we develop panel flow matching (PFM), a generative framework for learning distributions of
longitudinal data with irregular and sparse observation times. By estimating a parsimonious vector field
that transports a simple latent distribution to the observed panel distribution, the proposed method
provides a unified approach to panel density estimation, longitudinal data generation/completion, and
classification. A key advantage of PFM is that it avoids a preliminary dimension-reduction step for handling data irregularity and sparsity, and instead models the distributional structure through a flow-based latent model directly at the data level.
This framework provides a continuous representation of densities and data generators for longitudinal samples, making it applicable to data that exhibit complex nonlinear temporal variation or distributional features.

More broadly, flow-based methods can be viewed as a nonparametric approach to latent statistical modeling, where the infinitesimal and invertible nature of the flow connects the latent and observed spaces continuously and reversibly, providing flexible expressive power for the forward and backward mappings between the latent and observed spaces.
Beyond the longitudinal setting considered in this work, flow-based methods may also be useful for other types of complex data, such as manifold data \citep{chen2023flow} and stochastic process data \citep{li2014functional}, where latent structural modeling is often essential. We plan to investigate these directions in future research.
\end{sloppypar}

\bibliographystyle{apalike}
\bibliography{refbib}

\end{document}